\documentclass[preprint]{aastex61}
\usepackage{color} 
\shorttitle{He, C and N in Planetary Nebulae}
\shortauthors{Henry et al.}
\received{2017 June 9}
\revised{2017 July 28}
\accepted{2017 August 31}
\submitjournal{MNRAS}

\begin{document}

\title{On the Production of He, C and N by Low and Intermediate Mass Stars: A Comparison of Observed and Model-Predicted Planetary Nebula Abundances}


\author{R.B.C. Henry}
\affiliation{Department of Physics and Astronomy, University of Oklahoma, Norman, OK 73019, USA}

\author{B.G. Stephenson}
\affiliation{Department of Physics and Astronomy, University of Oklahoma, Norman, OK 73019, USA}

\author{M.M. Miller Bertolami}
\affiliation{Instituto de Astrof\'isica de La Plata, UNLP-CONICET, Paseo del Bosque s/n, B1900FWA, La Plata, Argentina}

\author{K.B. Kwitter} 
\affiliation{Department of Astronomy, Williams College, Williamstown, MA  01267, USA}
 
\author{B. Balick}
\affiliation{Department of Astronomy, University of Washington, Seattle, WA 98195, USA}

\correspondingauthor{R.B.C. Henry}

\email{rhenry@ou.edu}

\begin{abstract}

The primary goal of this paper is to make a direct comparison between the measured  and model-predicted abundances of He, C and N in a sample of 35 well-observed Galactic planetary nebulae (PN). All observations, data reductions, and abundance determinations were performed in house to ensure maximum homogeneity. Progenitor star masses ($M\le4~M_{\odot}$) were inferred using two published sets of post-AGB model tracks and L and T$_{eff}$ values. We conclude the following: 1)  the mean values of N/O across the progenitor mass range exceeds the solar value, indicating significant N enrichment in the majority of our objects; 2) the onset of hot bottom burning appears to begin around 2~M$_{\odot}$, i.e., lower than $\sim5~M_{\odot}$ implied by theory; 3) most of our objects show a clear He enrichment, as expected from dredge-up episodes; 4) the average sample C/O value is 1.23, consistent with the effects of third dredge-up; and 5) model grids used to compare to observations successfully span the distribution over metallicity space of all C/O and many He/H data points but mostly fail to do so in the case of N/O. The evident enrichment of N in PN and the general discrepancy between the observed and model-predicted N/O abundance ratios signal the need for extra-mixing as an effect of rotation and/or thermohaline mixing in the models. The unexpectedly high N enrichment that is implied here for low mass stars, if confirmed, will likely impact our conclusions about the source of N in the Universe.

\end{abstract}


\section{Introduction}

Galaxies evolve chemically because hydrogen-rich interstellar material forms stars which subsequently convert a fraction of the hydrogen into heavier elements. These nuclear products are expelled into the interstellar medium and thereby enrich it.  As this cycle is continuously repeated, the mass fraction of metals rises. Additional factors which influence the metal abundances in galaxies include the exchange of gas with the intergalactic medium via inflow and outflow. 

A crucial component for understanding the rate at which the interstellar abundance of a specific element rises over time is the amount of the element that is synthesized and expelled by a star of a specific mass during its lifetime, i.e., the stellar yield. Generally, stellar yields are estimated by computing stellar evolution models that predict them. These models are constrained using elemental abundance measurements of the material that is cast off from the star in the form of winds propelled by radiation pressure, periodic expulsions by stellar pulsations, or sudden ejection caused by explosions.

In the current study, we are interested in the production of He, C and N by low and intermediate mass stars (LIMS), that is, those stars typically considered to occupy the mass range of 1-8~M$_{\odot}$. Stellar models suggest that internal temperatures become sufficiently high either in the cores or outer shells of these stars to drive not only the conversion of H to He via the proton-proton chain reactions, but also the triple alpha process as well as the CN(O) cycle to produce C and N, respectively. Observationally, there is overwhelming evidence that LIMS do indeed synthesize and eventually expel measurable amounts of elements such as He, C, N and perhaps O, as well as s-process elements [see articles by \citet{herwig05}, \citet{kwitter12}, \citet{karakas14}, \citet{delgado16}, \citet{maciel17} and \citet{sterling17}]. However, the impact that LIMS actually have, relative to massive stars on the chemical evolution of these elements in a galaxy, is still very much open for debate. 

The material that is cast off by LIMS during and after the AGB stage in the form of winds of varying speeds can subsequently form large-scale density enhancements and become photoionized by the UV photons produced by the hot, shrinking stellar remnant, forming a planetary nebula (PN). The photon energy absorbed by the nebula results in the production of detectable emission lines that can be analyzed in detail to infer abundance, temperature and density information about the PN. 

PN abundance patterns reflect the nature of the chemical composition of the LIMS atmospheres at the end of stellar evolution and are therefore useful in two ways. First, the abundances of alpha, Fe-peak and r-process elements relative to H, especially O/H, Ne/H, S/H, Ar/H and Cl/H in PN, evidently represent the levels of these elements that were present in the interstellar material out of which the progenitor star formed. This conclusion is strongly supported by a recent study by \citet{maciel17}. This team has recently compiled and analyzed a database containing abundance measurements of 1318 PN along with a second database containing similar information about 936 H~II regions, the latter objects representing the current ISM abundance picture. Through the use of histograms and scatter plots, the authors show that both object types exhibit the same lockstep behavior of Ne/H, S/H and Ar/H, all versus O/H\footnote{The alpha elements O, Ne, S and Ar are apparently forged in massive stars by similar nuclear processes which transcend position and environment. Therefore, their relative abundances track each other.}. This familiar result strongly supports the idea that LIMS do not themselves alter the levels of the alpha elements that were present in the interstellar material out of which they formed. As a result, PN can be used as probes of ISM conditions at the time of progenitor star formation\footnote{We qualify this seemingly tidy picture by pointing out that oxygen enrichment in PN has been reported by \citet{pequignot00} and more recently in C-rich PN by \citet{delgado15} and \citet{garcia16}.}. 

\begin{figure}
   \includegraphics[width=6in,angle=270]{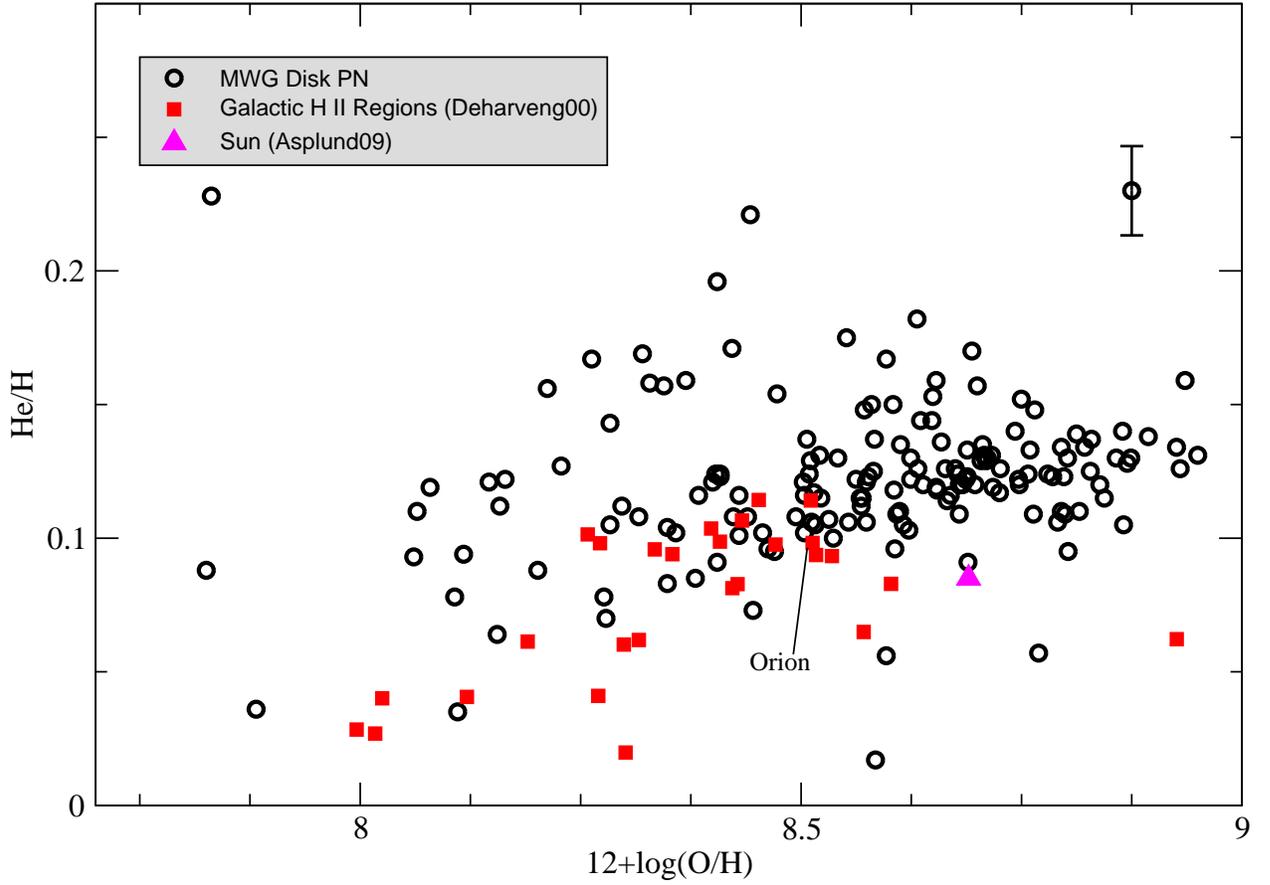} 
   \caption{He/H versus 12+log(O/H). Open black circles refer to PN located in the MWG disk and taken from our extended sample, filled red squares represent Galactic H~II regions from \citet{deharveng00} and the filled magenta triangle shows the solar position \citep{asplund09}. The position of Orion \citep{esteban04} is indicated.}
\label{he2hvo2h_BIG}
\end{figure}

\begin{figure}
   \includegraphics[width=6in,angle=270]{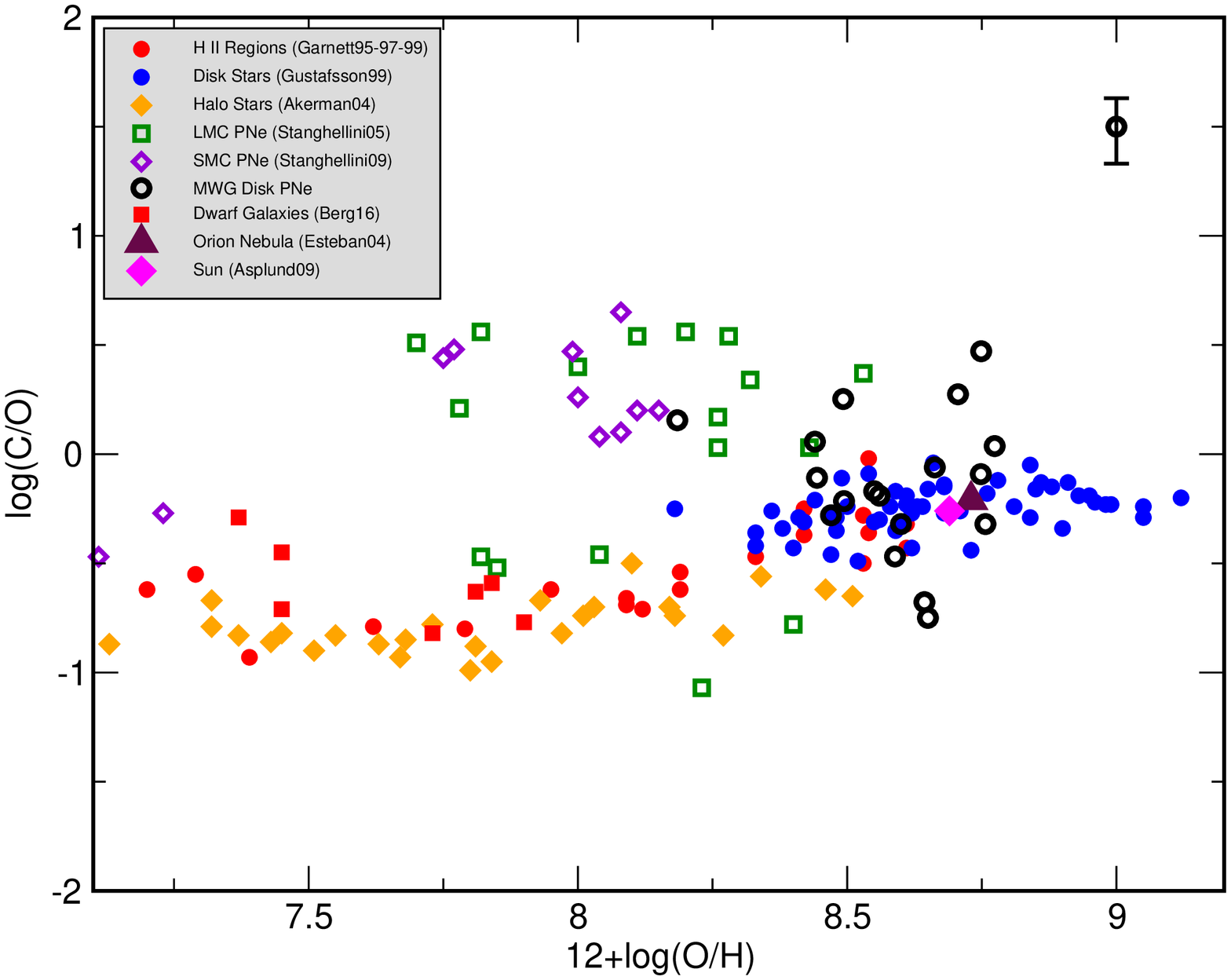} 
   \caption{log(C/O) versus 12+log(O/H). Open black circles refer to PN located in the MWG disk and taken from our extended sample, red filled circles represent H~II regions from Garnett (1995, 1997, 1999), MWG disk stars from \citet{gustafsson99} are shown with blue filled circles, MWG metal poor halo stars from \citet{akerman04} are indicated with orange filled diamonds, green open squares and diamonds indicated LMC and SMC PN by \citet{stanghellini05,stanghellini09} respectively, and red filled squares correspond to low metallicity dwarf galaxies by \citet{berg16}. The maroon filled triangle and magenta filled diamond represent Orion \citep{esteban04} and the Sun \citep{asplund09}, respectively.}
\label{c2ovo2h_BIG}
\end{figure}

\begin{figure}
   \includegraphics[width=6in,angle=270]{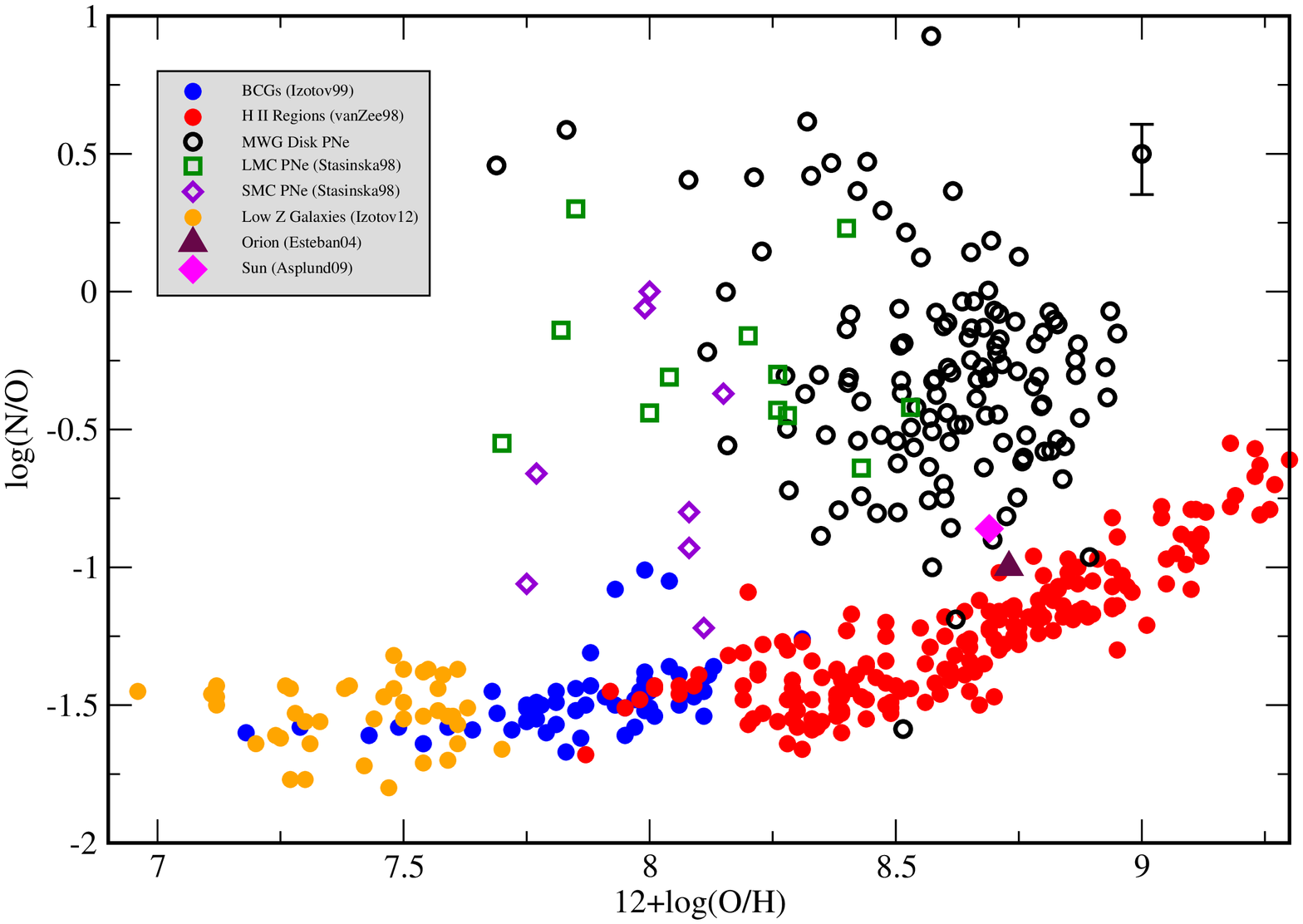} 
   \caption{log(N/O) versus 12+log(O/H). Open black circles refer to PN located in the MWG disk and taken from our extended sample, filled blue circles represent blue compact galaxies \citep{izotov99}, filled red circles are H~II regions \citep{vanzee98}, open green squares and open maroon diamonds are PN from the LMC and SMC, respectively, from \citet{stasinska98}, and filled orange circles are low metallicity galaxies from \citet{izotov12}. The maroon filled triangle and magenta filled diamond represent Orion \citep{esteban04} and the Sun \citep{asplund09}, respectively. }
\label{n2ovo2h_BIG}
\end{figure}

Second, and more relevant to our current study, elements such as He, C, N and s-process elements are found to be enriched in PN, and so measurements of their abundances provide valuable information about the nucleosynthesis that occurs during the lifetime of PN progenitor stars. Figures~\ref{he2hvo2h_BIG}, \ref{c2ovo2h_BIG} and \ref{n2ovo2h_BIG} are plots showing He/H, log(C/O) and log(N/O), respectively, versus 12+log(O/H), where O/H is taken as the gauge of overall metallicity. Each plot contrasts the values for PN (open symbols) with analogous values of objects such as H~II regions and F and G dwarfs, all of which measure the interstellar values of the two ratios involved either currently (H~II regions) or at the time of their formation (stars). Original data for the MWG Disk PN points in these three figures can be found in \citet{henry00,henry04,milingo10,kwitter12,dufour15}.

The relatively narrow horizontal band (especially in the cases of C/O and N/O) populated by the H~II regions and stars in each graph demonstrates how He/H, C/O and N/O generally behave as metallicity changes. These patterns of chemical evolution are reflections of the details of stellar evolution and nucleosynthesis, processes which apparently are universal and space invariant. Presumably, when the progenitor stars of the PN in the plots began their lives on the main sequence, they were located along these bands at a position near the PN's current O/H value. 

PN values of He/H, C/O and N/O clearly fall above these bands in nearly every case, strongly suggesting that {\it He, C, and N have been significantly enriched by nucleosynthesis in nearly all progenitor stars over their lifetimes}. 
High PN values for these three abundance ratios have been observed previously. For example, \citet{henry90} compiled the He/H and log(N/O) measurements by \citet{aller83,aller87} for 84 Galactic PN and found the average values of these two ratios to be 0.11 and -0.38, respectively. From their large sample of southern PN, \citet{kb94} found similar average values for He/H and log(N/O) of 0.115 and -0.33, respectively. In the case of C/O, the log of our average value for objects in the current study (see Table~\ref{abuns}) is log(C/O)=0.088 compared with 0.06 from \citet{kb94} (see their Table~14)\footnote{Note that we have estimated their averages for N/O and C/O from their separate averages of N, C and O.}. In addition, simple eyeball comparisons of the ranges of all three ratios shown in \citet{henry90} with \citet[erratum]{henry90err}, the figures in \citet{kb94} and our Figs.~\ref{he2hvo2h_BIG}-\ref{n2ovo2h_BIG} in the current paper show good consistency among these studies and reinforce the point that these ratios in PN are generally enhanced relative to levels of found in H~II regions of similar metallicity.

Regarding the apparent N enhancement in PN in particular, theory predicts that the hot bottom burning process that explains the extent of the enrichment occurs in the AGB stage of stars whose progenitors were at least 3-4~M$_{\odot}$ depending upon the star's metallicity. Yet, based upon the properties of the stellar initial mass function, we also know that in the absence of an unknown selection effect, most of the PN included in these figures must be the products of relatively low mass progenitors, i.e., 1-2~M$_{\odot}$ and should therefore show very little N enrichment. How can we reconcile this observational result with theory? 

{\it The purpose of our investigation here is to confront recently published stellar model predictions of PN abundances with the observed abundances of He, C, N and O.} We consider the models of four different research groups and evaluate each set of models based upon how well they appear to explain the observed abundances. Since these same models also predict the total stellar yield of each element, only a fraction of which is present in the visible nebula, our results can be used to assess the relevance of the yield predictions for use in chemical evolution models.

Previous studies comparing model predictions and observations have been carried out by \citet{marigo03,marigo11}, \citet{stanghellini09}, \citet{delgado15}, \citet{ventura15}, \citet{lugaro16}, and \citet{garcia16}. The principle method of comparison for these studies features plots of two different element-to-element ratios, e.g., C/O versus N/O, showing both the observed abundances and model tracks computed for a range of stellar masses. Most authors find that abundance trends involving He, C and N can be explained by various amounts of 3rd dredge-up, which elevates C, and hot bottom burning, which does likewise to N. However, explanations of PN abundance patterns based upon progenitor masses is typically not included. 

Our study augments earlier analyses by also considering each of the ratios of He/H, C/O or N/O separately {\it as a function of an object's progenitor mass}. The sample of PN abundances which we compare to model predictions consists of 35 objects that have previously been observed and analyzed by our group. We have observed all objects in the optical with ground-based telescopes and 13 out of the 35 PN in the UV using either IUE or HST.

We describe the PN sample in detail in section~2. Our methods for determining the necessary abundances and progenitor mass for each object are provided in section~3. A description of each stellar modelling code used to predict the PN abundances and stellar yields of He, C and N, along with an analysis of our comparison of theory and observation are presented in section~4. Our summary and conclusions appear in section~5.

\section{Object Sample}

For nearly 25 years our team has been building a spectroscopic database comprising 166 planetary nebulae located primarily in the disk and halo of the Milky Way Galaxy. While a vast majority of the observations have necessarily been restricted to the optical region of the spectrum, i.e., 3700\AA~to 10,000\AA, we have also collected UV data for a smaller sample using both the IUE and HST facilities. Most of these data, along with derived abundances of He, N, O, Ne, S, Ar, and in several cases C, have been published. Because we are currently interested in comparing our observed CNO abundances in PN with theoretical predictions of the abundances of these same elements as a function of initial stellar mass, it is necessary to identify a subset of our database for which we can infer progenitor star masses that are based upon carefully and consistently determined central star luminosities and effective temperatures.

Initial stellar masses can be derived by using published values for T$_{eff}$ and log(L/L$_{\odot}$) of each central star to place the star in a theoretical HR diagram. After plotting post-AGB evolutionary tracks labeled by mass in the same diagram, stellar masses can be inferred by interpolating between tracks\footnote{We are very much aware of the pitfalls of using this method to determine central star and progenitor star masses. Problems stem primarily from the small separation between adjacent model evolutionary tracks in the luminosity-temperature plane that are used to infer these masses, given the uncertainties of the observed values of these two parameters. However, we are confident that in using this method we can at least tell if a progenitor star is inside or outside of a mass range for which theory predicts C enrichment through triple alpha burning and dredge-up, or N enrichment through hot bottom burning.}. 

The extensive compilation of stellar data by \citet[Tables 9.5 and 9.6, each comprising 210 objects]{frew08} was adopted as our source of T$_{eff}$ and log(L/L$_{\odot}$) for reasons of consistency. A total of 32 objects with N and O abundances from our database were also listed in the Frew paper. We have also measured C abundances using UV emission lines of C~III] $\lambda\lambda$1907,1909 for 10 of the 32 PN. Besides abundances of N and O, we have determined C abundances for three other objects in our database which are not part of the Frew list and have included these objects in order to maximize the sample size for objects with measured C abundances. Thus, our final object list contains 35 PN (about 1/5 of the objects in our original database), all of which have measured N and O abundances and including 13 objects with measured C abundances. We emphasize the fact that the spectroscopic observations of the 35 PN, as well as the data reductions and abundance analyses, were carried out exclusively by members of our team.   

Our final sample of 35 objects is listed in Table~\ref{objects}\footnote{Fg1 and NGC~6826 are the only objects in our sample with any evidence of binary central stars. According to  \citet{boffin12}, Fg1 has a period of 1.2d. NGC~6826 has a fast rotating central star, which is something that can only be achieved in a merger \citep{demarco15}. However, neither of these objects exhibits any abundance peculiarities, according to our data. For now, we have assumed that the presence of a secondary star does not affect our results.}. For each PN identified in column~1 we provide a morphological description in column~2 and the Peimbert type in column~3. Column~4 indicates the spectral range over which we have observed the object (OP=optical; IUE/HST=UV data source). Finally, columns~5 and 6 list the galactocentric distance in kiloparsecs and the vertical height in parsecs above the Galactic plane for each object. Taking the distance of the Sun from the Galactic center as 8~kpc and the scale height of the thin disk as about 350~pc, we see that most of the PN in our sample are located near the solar neighborhood and within the thin disk. We also note that while the values of the He/H, C/O and N/O abundance ratios over the MW disk are sensitive to metallicity as measured by O/H, the O/H ratio only decreases by 0.23 dex between 6 and 10 kpc in galactocentric distance, assuming an O/H gradient of -0.058 dex/kpc \citep{henry10}. From Figs.~\ref{he2hvo2h_BIG}-\ref{n2ovo2h_BIG}, this corresponds to only minor changes in He/H, C/O and N/O, and so we can ignore the effects of the disk's metallicity gradient.

\begin{deluxetable}{cccccc}
\tablecolumns{6}
\tablewidth{0pc}
\tabletypesize{\scriptsize}
\tablecaption{Sample Objects\label{objects}}
\tablehead{
\colhead{PN}&{Morphology}&{Peimbert Type}&{Spectral Range\tablenotemark{a}}&{R$_G$(kpc)\tablenotemark{b}}&{z(pc)\tablenotemark{c}}
}
\startdata
FG1	&	elliptical; bipolar jets	&	II	&	OP		&	7.56	&	251.07	\\
IC2149	&	round/complex	&	II	&	OP		&	8.46	&	276.46	\\
IC2165	&	elliptical	&	II	&	HST,OP		&	9.98	&	-536.80	\\
IC3568	&	round	&	II	&	HST,OP		&	9.53	&	1642.47	\\
IC418	&	elliptical	&	II	&	IUE,OP		&	8.92	&	-493.40	\\
IC4593	&	elliptical	&	II	&	IUE,OP		&	6.94	&	1026.64	\\
N1501	&	elliptical	&	II	&	OP		&	8.59	&	82.12	\\
N2371	&	barrel	&	II	&	OP		&	9.31	&	478.51	\\
N2392	&	elliptical	&	II	&	IUE,OP		&	9.17	&	382.74	\\
N2438	&	round	&	II	&	OP		&	8.95	&	102.01	\\
N2440	&	pinched-waist/multisymmetric	&	I	&	HST,OP		&	9.23	&	80.22	\\
N2792	&	round	&	II	&	OP	&	8.39	&	144.41	\\
N3195	&	barrel	&	I	&	OP		&	7.35	&	-688.73	\\
N3211	&	elliptical	&	II	&	OP		&	7.69	&	-162.14	\\
N3242	&	elliptical/shells/ansae	&	II	&	HST,OP		&	8.18	&	530.62	\\
N3918	&	barrel	&	II	&	OP		&	7.42	&	151.08	\\
N5315	&	elliptical/multisymmetric	&	I	&	HST,OP		&	6.93	&	-80.02	\\
N5882	&	elliptical/shells	&	II	&	HST,OP		&	6.64	&	297.52	\\
N6369	&	round/shells	&	II	&	OP		&	6.46	&	157.97	\\
N6445	&	irregular/lobe-remnants	&	I	&	OP		&	6.63	&	94.78	\\
N6537	&	pinched-waist	&	I	&	OP		&	6.04	&	25.83	\\
N6563	&	elliptical/lobes	&	II	&	OP		&	6.34	&	-213.34	\\
N6567	&	unresolved	&	II	&	OP		&	6.36	&	-19.06	\\
N6572	&	elliptical/multisymmetric/lobes	&	I	&	OP		&	6.58	&	381.92	\\
N6629	&	elliptical/halo	&	II	&	OP		&	6.04	&	-176.04	\\
N6751	&	round/flocculent	&	I	&	OP		&	6.34	&	-206.96	\\
N6804	&	barrel/shell	&	II	&	OP		&	7.06	&	-117.63	\\
N6826	&	elliptical/shell/halo/ansae	&	II	&	IUE,OP		&	7.96	&	287.77	\\
N6894	&	round	&	I	&	OP		&	7.64	&	-59.88	\\
N7008	&	elliptical	&	II	&	OP		&	8.07	&	66.97	\\
N7009	&	elliptical/shell/halo/ansae	&	II	&	IUE,OP	&	7.09	&	-822.70	\\
N7027	&	elliptical/multisymmetric/hourglass-shell/halo	&	II	&	OP		&	7.97	&	-51.89	\\
N7293	&	round/shells/bowshocks	&	I	&	IUE,OP		&	7.90	&	-184.75	\\
N7354	&	barrel/shell/jets	&	I	&	OP		&	8.62	&	64.76	\\
N7662	&	elliptical/shell/halo	&	II	&	HST,OP		&	8.42	&	-380.96	\\
\enddata
\tablenotetext{a}{HST=Hubble Space Telescope; IUE=International Ultraviolet Explorer; OP=Ground-based optical telescopes. HST observations spanned the UV and optical spectral regions, while the IUE covered the UV. The ground-based observations normally extended from 3700{\AA} to 1 micron.}
\tablenotetext{b}{$R_G=\{R_{\odot}^2-[cos(b) \times D]^2-2\times R_{\odot} \times D \times cos(l) \times cos(b)\}^{1/2}$, D is the object's heliocentric distance, $R_{\odot}$ is the Sun's galactocentric distance of 8~kpc, and $l$ and $b$ are heliocentric galactic coordinates. $D$, $l$ and $b$ are taken from \citet{frew08}.}
\tablenotetext{c}{$z=D\times sin(b)$, where $D$ and $b$ are the object's heliocentric distance and galactic longitude, respectively, and are taken from \citet{frew08}.}
\end{deluxetable}

Table~\ref{journal} provides the details concerning the observations of each of our 35 sample objects. The name of the PN appears in column~1. Columns~2-7 list the observation date, the telescope(s) and instrument(s) used, the times for the blue and red exposures, and the offset from the central star, respectively. The relevant references for the observations are given in column~8. 

Beginning with our first project in 1993, all data have been reduced and measured manually by one of us (KBK) using the same techniques throughout. Uncertainties were explicitly measured and calculated in our early papers; then experience taught us that we could estimate them from the lines strengths themselves. ELSA (see \S3.1) calculates statistical uncertainties, but no systematics are included. The former are then propagated through to the final intensities and diagnostics. Systematic errors are minimized by employing the same set of atomic data for abundance determinations throughout and by having a homogeneous data reduction and measuring pipeline, all performed by the same individual. The original line strengths are available in the relevant papers provided in Table~\ref{journal}.

\begin{deluxetable}{llllllll}
\tablecolumns{8}
\tabletypesize{\scriptsize}
\tablecaption{Journal Of Observations\label{journal}}
\tablehead{
\colhead{PN}&{Observation Date}&{Telescope}&{Instrument\tablenotemark{a}}&{Exp Time B}&{Exp Time R}&{Offset From CS}&{Ref\tablenotemark{b}}
}
\startdata
Fg1			&	March-April 1997	&	CTIO 1.5-m	&	Cass spec	&	1500	&	900	&	\nodata	&	1	\\
IC 2149			&	2007 Jan	&	APO 3.5m	&	DIS 	&	90	&	90	&	\nodata	&	2	\\
IC 2165			&	1996 Dec; HST Cy 19	&	KPNO 2.1m	&	 Goldcam	&	330	&	90	&	\nodata	&	3,4	\\
IC 3568			&	1996 May; HST Cy 19	&	KPNO 2.1m	&	 Goldcam	&	120	&	120	&	4"N	&	5,4	\\
IC 418			&	1996 Dec	&	KPNO 2.1m	&	 Goldcam	&	30	&	440	&	5"N	&	6	\\
IC 4593			&	1996 May;	&	KPNO 2.1m	&	 Goldcam	&	180	&	600	&	3"S	&	5	\\
NGC 1501			&	2007 Jan	&	APO 3.5m	&	DIS	&	240	&	240	&	\nodata	&	2	\\
NGC 2371			&	1996 Dec	&	KPNO 2.1m	&	 Goldcam	&	300	&	300	&	9.7"S, 15.7"W	&	3	\\
NGC 2392			&	1996 Dec	&	KPNO 2.1m	&	 Goldcam	&	450	&	2520	&	14"S	&	6	\\
NGC 2438			&	1996 Dec	&	KPNO 2.1m	&	 Goldcam	&	300	&	300	&	16.3"N	&	3	\\
NGC 2440			&	1996 Dec	&	KPNO 2.1m	&	 Goldcam	&	100	&	360	&	4"S	&	3,4	\\
NGC 2792			&	March-April 1997	&	CTIO 1.5-m	&	Cass spec&	2400	&	300	&	\nodata	&	1	\\
NGC 3195			&	March-April 1997	&	CTIO 1.5-m	&	Cass spec	&	1200	&	900	&	\nodata	&	1	\\
NGC 3211			&	March-April 1997	&	CTIO 1.5-m	&	Cass spec	&	480	&	600	&	\nodata	&	1	\\
NGC 3242			&	1996 Dec	&	KPNO 2.1m	&	 Goldcam	&	450	&	480	&	8"S	&	1,4	\\
NGC 3918			&	March-April 1997	&	CTIO 1.5-m	&	Cass spec 	&	100	&	720	&	\nodata	&	3	\\
NGC 5315			&	 2004 August; HST Cy 19	&	CTIO 1.5-m	&	Cass spec 	&	961	&	855	&	\nodata	&	7,4	\\
NGC 5882			&	March-April 1997; HST Cy 19	&	CTIO 1.5-m	&	Cass spec 	&	390	&	480	&	\nodata	&	3,4	\\
NGC 6369			&	2003 June	&	KPNO 2.1m	&	 Goldcam	&	2000	&	2200	&	10"N	&	7	\\
NGC 6445			&	2003 June	&	KPNO 2.1m	&	 Goldcam	&	1200	&	1200	&	25"N	&	7	\\
NGC 6537			&	2003 June	&	KPNO 2.1m	&	 Goldcam	&	725	&	300	&	\nodata	&	7	\\
NGC 6563			&	March-April 1997	&	CTIO 1.5-m	&	Cass spec 	&	1200	&	600	&	\nodata	&	1	\\
NGC 6567			&	March-April 1997	&	CTIO 1.5-m	&	Cass spec	&	390	&	330	&	\nodata	&	3	\\
NGC 6572			&	1999 June	&	KPNO 2.1m	&	 Goldcam	&	72	&	72	&	\nodata	&	8	\\
NGC 6629			&	March-April 1997	&	CTIO 1.5-m	&	Cass spec 	&	420	&	360	&	\nodata	&	1	\\
NGC 6751			&	2003 June	&	KPNO 2.1m	&	 Goldcam	&	1500	&	1500	&	6"S	&	7	\\
NGC 6804			&	2003 June	&	KPNO 2.1m	&	 Goldcam	&	1800	&	5100	&	10"S	&	7	\\
NGC 6826			&	1996 May;	&	KPNO 2.1m	&	 Goldcam	&	240	&	720	&	9"S	&	5	\\
NGC 6894			&	1999 June	&	KPNO 2.1m	&	 Goldcam	&	600	&	960	&	\nodata	&	8	\\
NGC 7008			&	2004 August	&	KPNO 2.1m	&	 Goldcam	&	1200	&	1926	&	29"N,11"E	&	7	\\
NGC 7009			&	1996 May;	&	KPNO 2.1m	&	 Goldcam	&	90	&	60	&	9"S	&	5	\\
NGC 7027			&	1996 May	&	KPNO 2.1m	&	 Goldcam	&	25	&	110	&	\nodata	&	3	\\
NGC 7293			&	1996 Dec	&	KPNO 2.1m	&	 Goldcam	&	1800	&	1800	&	 97"E, 171"N	&	9	\\
NGC 7354			&	2003 June	&	KPNO 2.1m	&	 Goldcam	&	3503	&	4500	&	\nodata	&	7	\\
NGC 7662			&	1999 June; HST Cy 19	&	KPNO 2.1m	&	 Goldcam	&	90	&	100	&	\nodata	&	3,4	\\
\enddata
\tablenotetext{a}{Slit dimensions width x length in arcseconds (length oriented E-W)-- Cass Spectrograph: 5x320; DIS: 2x360; Goldcam: 5x285}
\tablenotetext{b}{References-- 1: \citet{milingo02};  2: \citet{henry10};  3: \citet{kwitter03}; 4: \citet{dufour15};  5: \citet{kwitter98};  6: \citet{henry00}; 7: \citet{milingo10};  8: \citet{kwitter01};  9: \citet{henry99}}
\end{deluxetable}


\section{Methods}

\subsection{Nebular Abundances}

We have published abundances of He, N, O and in some cases C previously in papers indicated in the footnote to column~8 in Table~\ref{journal}. However, we sought to render the abundances more homogeneous by recomputing all of them using the same updated abundance code along with the newly-published ionization correction factors by \citet{delgado14} in the cases of total He, C and O abundances.

Ionic abundances were determined using the code ELSA (Emission Line Spectral Analysis), a program whose core is a 5-level atom routine. Emission line strengths and their uncertainties used as input to ELSA were taken from the references listed in Table~\ref{journal}. We used an updated version of the program originally introduced by \citet{johnson06}, where the major change was the addition of a C~III] density diagnostic routine based upon the $\lambda$1907$/\lambda$1909 line strength ratio (C~III] $\lambda$1909 was already included in the program). The important emission lines besides H$\beta$ that were used in the ionic abundance computations for each object were He~I $\lambda$5876, He~II $\lambda$4686, C~III] $\lambda\lambda$1907,1909, [N~II] $\lambda$6584, [O~II] $\lambda$3727, [O~III] $\lambda$5007 and [O~III] $\lambda$4363.

The resulting ionic abundances and uncertainties with respect to H$^+$ produced by ELSA are presented in Table~\ref{ions}. The object names are given in column~1 followed by column pairs containing the abundances and uncertainties for each ion labeled in the header. Uncertainties for the ionic abundances are computed internally by ELSA and are the result of contributions from:  1)~ the uncertainties in the line strength ratios, e.g., I$_{\lambda}$/I$_{H\beta}$; and 2)~ the uncertainties in the reaction rate coefficients (radiative recombination or collisional excitation rate coefficients) that stem from errors in electron temperature.

\movetabledown=3.0 in
\begin{deluxetable}{lcccccccccccc}
\tabletypesize{\scriptsize}
\tablecolumns{13}
\tablewidth{0pc}
\tablecaption{Ionic Abundance\label{ions}}
\tablehead{
\colhead{PN} & {He$^+$} & {$\sigma$He$^+$} & {He$^{+2}$} & {$\sigma$He$^{+2}$} & {C$^{+2}$} & {$\sigma$C$^{+2}$} & {N$^+$} & {$\sigma$N$^+$} & {O$^+$} & {$\sigma$O$^+$} & {O$^{+2}$} & {$\sigma$O$^{+2}$}
}
\startdata
FG1	&	1.16E-01	&	1.26E-02	&	1.37E-02	&	1.88E-03	&	\nodata	&	\nodata	&	1.15E-05	&	1.40E-06	&	1.90E-05	&	5.50E-06	&	2.69E-04	&	2.45E-05	\\
IC2149	&	1.05E-01	&	1.20E-02	&	8.32E-05	&	4.03E-05	&	\nodata	&	\nodata	&	5.87E-06	&	8.00E-07	&	4.14E-05	&	9.80E-06	&	1.79E-04	&	1.72E-05	\\
IC2165	&	5.70E-02	&	7.09E-03	&	4.90E-02	&	5.15E-03	&	1.72E-04	&	1.30E-05	&	4.61E-06	&	1.00E-07	&	1.27E-05	&	2.00E-06	&	1.36E-04	&	5.00E-06	\\
IC3568	&	1.14E-01	&	1.26E-02	&	3.42E-03	&	1.33E-04	&	1.55E-04	&	2.60E-05	&	3.92E-07	&	8.10E-08	&	8.85E-06	&	5.46E-06	&	2.82E-04	&	1.70E-05	\\
IC418	&	7.00E-02	&	8.26E-03	&	\nodata	&	\nodata	&	3.58E-04	&	2.03E-04	&	3.78E-05	&	7.00E-06	&	8.38E-05	&	2.11E-05	&	6.90E-05	&	6.00E-06	\\
IC4593	&	1.02E-01	&	1.18E-02	&	5.40E-04	&	8.30E-05	&	8.54E-04	&	1.21E-04	&	3.45E-06	&	9.00E-07	&	4.38E-05	&	2.50E-05	&	3.71E-04	&	3.01E-05	\\
N1501	&	8.56E-02	&	9.30E-03	&	3.86E-02	&	5.30E-03	&	\nodata	&	\nodata	&	1.95E-06	&	2.00E-07	&	9.89E-06	&	1.94E-06	&	3.20E-04	&	2.56E-05	\\
N2371	&	2.62E-02	&	2.93E-03	&	8.03E-02	&	1.13E-02	&	\nodata	&	\nodata	&	9.37E-06	&	2.25E-06	&	1.66E-05	&	9.60E-06	&	1.36E-04	&	1.66E-05	\\
N2392	&	5.80E-02	&	6.50E-03	&	2.10E-02	&	2.91E-03	&	1.26E-04	&	5.90E-05	&	1.37E-06	&	1.32E-06	&	2.53E-05	&	7.50E-06	&	1.14E-04	&	1.60E-05	\\
N2438	&	7.53E-02	&	1.95E-02	&	2.08E-02	&	2.82E-03	&	\nodata	&	\nodata	&	2.95E-05	&	5.90E-06	&	5.04E-05	&	2.81E-05	&	2.72E-04	&	2.69E-05	\\
N2440	&	5.35E-02	&	6.89E-03	&	7.39E-02	&	6.72E-03	&	1.42E-04	&	1.60E-05	&	7.46E-05	&	8.00E-06	&	4.38E-05	&	1.50E-05	&	1.25E-04	&	7.00E-06	\\
N2792	&	1.92E-02	&	2.54E-03	&	9.16E-02	&	1.33E-02	&	\nodata	&	\nodata	&	3.92E-07	&	2.96E-07	&	1.64E-06	&	9.90E-07	&	1.11E-04	&	1.61E-05	\\
N3195	&	1.24E-01	&	1.35E-02	&	1.16E-02	&	1.62E-03	&	\nodata	&	\nodata	&	9.59E-05	&	1.17E-05	&	1.49E-04	&	4.50E-05	&	3.06E-04	&	2.66E-05	\\
N3211	&	3.13E-02	&	3.71E-03	&	8.39E-02	&	1.21E-02	&	\nodata	&	\nodata	&	1.22E-06	&	4.60E-07	&	4.39E-06	&	4.21E-06	&	1.85E-04	&	2.58E-05	\\
N3242	&	6.88E-02	&	1.09E-02	&	4.58E-02	&	3.77E-03	&	1.65E-04	&	8.00E-06	&	5.72E-07	&	5.40E-08	&	5.42E-06	&	1.46E-06	&	2.40E-04	&	5.00E-06	\\
N3918	&	7.02E-02	&	8.78E-03	&	4.43E-02	&	6.12E-03	&	\nodata	&	\nodata	&	1.00E-05	&	1.70E-06	&	1.98E-05	&	6.00E-06	&	2.62E-04	&	3.24E-05	\\
N5315	&	1.32E-01	&	1.59E-02	&	\nodata	&	\nodata	&	2.22E-04	&	2.70E-05	&	1.91E-05	&	1.70E-06	&	1.21E-05	&	3.30E-06	&	3.47E-04	&	1.70E-05	\\
N5882	&	1.03E-01	&	1.33E-02	&	6.92E-03	&	4.65E-04	&	7.83E-05	&	2.06E-05	&	1.81E-06	&	5.00E-08	&	5.34E-06	&	8.90E-07	&	4.11E-04	&	3.10E-05	\\
N6369	&	1.30E-01	&	1.48E-02	&	1.65E-03	&	2.24E-04	&	\nodata	&	\nodata	&	1.11E-05	&	1.40E-06	&	2.22E-05	&	5.35E-06	&	4.72E-04	&	6.78E-05	\\
N6445	&	9.73E-02	&	1.05E-02	&	4.02E-02	&	5.50E-03	&	\nodata	&	\nodata	&	8.72E-05	&	1.28E-05	&	1.01E-04	&	3.40E-05	&	3.62E-04	&	3.92E-05	\\
N6537	&	9.60E-02	&	1.44E-02	&	7.43E-02	&	1.08E-02	&	\nodata	&	\nodata	&	2.89E-05	&	5.90E-06	&	3.25E-06	&	1.08E-06	&	1.10E-04	&	1.72E-05	\\
N6563	&	1.11E-01	&	1.16E-02	&	1.50E-02	&	2.04E-03	&	\nodata	&	\nodata	&	5.35E-05	&	7.40E-06	&	1.22E-04	&	4.30E-05	&	2.80E-04	&	2.81E-05	\\
N6567	&	1.01E-01	&	1.44E-02	&	1.37E-03	&	1.08E-02	&	\nodata	&	\nodata	&	1.50E-06	&	3.30E-07	&	4.80E-06	&	1.67E-06	&	2.22E-04	&	2.38E-05	\\
N6572	&	1.25E-01	&	1.50E-02	&	5.64E-04	&	1.64E-04	&	\nodata	&	\nodata	&	6.89E-06	&	1.19E-06	&	7.27E-06	&	1.70E-06	&	3.72E-04	&	3.63E-05	\\
N6629	&	1.09E-01	&	1.24E-02	&	1.03E-03	&	3.15E-04	&	\nodata	&	\nodata	&	2.33E-06	&	5.50E-07	&	1.54E-05	&	7.40E-06	&	3.93E-04	&	3.30E-05	\\
N6751	&	1.36E-01	&	1.51E-02	&	\nodata	&	\nodata	&	\nodata	&	\nodata	&	4.68E-05	&	6.20E-06	&	7.93E-05	&	2.14E-05	&	3.17E-04	&	3.00E-05	\\
N6804	&	2.10E-02	&	2.67E-03	&	8.86E-02	&	1.27E-02	&	\nodata	&	\nodata	&	1.38E-07	&	9.30E-08	&	1.04E-06	&	2.70E-07	&	1.09E-04	&	1.46E-05	\\
N6826	&	1.07E-01	&	1.38E-02	&	\nodata	&	\nodata	&	3.93E-04	&	1.18E-04	&	2.01E-06	&	5.50E-06	&	1.67E-05	&	9.70E-06	&	3.59E-04	&	3.20E-05	\\
N6894	&	1.14E-01	&	1.22E-02	&	1.55E-02	&	2.12E-03	&	\nodata	&	\nodata	&	7.37E-05	&	9.60E-06	&	9.36E-05	&	2.95E-05	&	2.61E-04	&	3.91E-05	\\
N7008	&	7.80E-02	&	8.39E-03	&	7.02E-02	&	9.79E-03	&	\nodata	&	\nodata	&	1.22E-06	&	2.40E-07	&	2.15E-06	&	1.09E-06	&	3.02E-04	&	3.07E-05	\\
N7009	&	1.10E-01	&	1.26E-02	&	9.43E-03	&	9.79E-03	&	6.89E-04	&	4.31E-04	&	8.46E-07	&	1.09E-07	&	2.05E-06	&	4.60E-07	&	4.85E-04	&	4.50E-05	\\
N7027	&	6.16E-02	&	9.36E-03	&	4.35E-02	&	6.11E-03	&	\nodata	&	\nodata	&	6.79E-06	&	1.05E-06	&	6.65E-06	&	8.10E-07	&	1.84E-04	&	2.50E-05	\\
N7293	&	1.12E-01	&	1.41E-02	&	7.99E-03	&	4.07E-04	&	1.84E-04	&	3.35E-04	&	5.52E-05	&	7.42E-06	&	7.50E-05	&	5.90E-05	&	3.40E-04	&	3.60E-05	\\
N7354	&	9.02E-02	&	1.02E-02	&	3.96E-02	&	5.44E-03	&	\nodata	&	\nodata	&	7.82E-06	&	1.11E-06	&	6.38E-06	&	1.84E-06	&	3.47E-04	&	3.56E-05	\\
N7662	&	6.72E-02	&	6.18E-03	&	5.47E-02	&	7.91E-03	&	1.28E-04	&	1.40E-05	&	5.68E-07	&	3.50E-08	&	4.64E-06	&	1.20E-06	&	1.95E-04	&	1.00E-05	\\
\enddata
\end{deluxetable}

Ionic abundances in Table~\ref{ions} were converted to the total elemental abundance ratios of interest here, i.e., He/H, C/O, N/O, and O/H, by multiplying the value of (He$^{+2}$ + He$^+$)/H$^+$, C$^{+2}$/O$^{+2}$, N$^+$/O$^+$ and (O$^{+2}$ + O$^+$)/H$^+$, respectively, by a relevant ionization correction factor (ICF). Except in the case of N/O, ICFs and their uncertainties were determined using the schemes of \citet{delgado14}. The ICF value for He/H was taken as unity for each object, since negligible amounts of neutral He are expected to be present in PN (see the next paragraph). On the other hand, the values for the C/O and O/H ICFs along with their uncertainties are different for each object and are therefore provided in Table~\ref{icfs}. For N/O we followed \citet{kb94} and \citet{kwitter01} and assumed that N/O=(N$^+$/O$^+$).\footnote{Since the radiation- or density-bounded natures of our PN are unknown, Delgado-Inglada (private communication) recommended that we use this ICF instead of the one published in \citet{delgado14}.}

We have assumed throughout that the contribution of neutral He is negligible in all objects. With the possible exception of IC~418, this is justified by the fact that the O$^{+2}$/O$^+$ abundance ratio is greater than unity (see Table~\ref{ions}), since the ionization potential of O$^+$ (35.1eV) greatly exceeds that of He$^o$ (24.6eV). Concerning IC~418, \citet{dopita17} recently published the results of new high resolution integral field spectroscopy for this PN. Their observations show both moderate [O~II] $\lambda$3727 and [O~III] $\lambda$5007 strengths, weak [O~I] $\lambda$6300 and no He~I $\lambda$4686, qualitatively similar to the findings in \citet{henry00} and \citet{sharpee03}.  \citet{dopita17} also construct a detailed nebular model that implies an abundance ratio of He/H=0.11, significantly higher than our value of 0.07. Therefore, our neglect of neutral He in IC~418 may be unwarranted, in which case our inferred He abundance may in fact be too low. This uncertainty obviously affects the position of IC~418, currently at He/H=0.07, in Figs.~\ref{he2h}, \ref{c2ovhe2h} and \ref{n2ovhe2h}.

\begin{deluxetable}{lcccc}
\tablecolumns{5}
\tablewidth{0pc}
\tabletypesize{\scriptsize}
\tablecaption{Ionization Correction Factors For C/O and O/H\tablenotemark{a}\label{icfs}}
\tablehead{
\colhead{PN}	&	{ICF(C/O)}	& {$\sigma$ICF(C/O)} &	{ICF(O/H)}	&	{$\sigma$ICF(O/H)}
}

\startdata
FG1	&	\nodata	&	\nodata	&	1.06	&	0.17	\\
IC2149	&	\nodata	&	\nodata	&	1.00	&	0.15	\\
IC2165	&	1.09	&	0.11	&	1.50	&	0.24	\\
IC3568	&	1.17	&	0.13	&	1.02	&	0.17	\\
IC418	&	0.64	&	0.04	&	1.00	&	0.11	\\
IC4593	&	1.06	&	0.45	&	1.00	&	0.16	\\
N1501	&	\nodata	&	\nodata	&	1.26	&	0.21	\\
N2371	&	\nodata	&	\nodata	&	2.93	&	0.46	\\
N2392	&	0.96	&	0.08	&	1.21	&	0.18	\\
N2438	&	\nodata	&	\nodata	&	1.15	&	0.18	\\
N2440	&	0.87	&	0.06	&	1.84	&	0.27	\\
N2792	&	\nodata	&	\nodata	&	3.99	&	0.66	\\
N3195	&	\nodata	&	\nodata	&	1.05	&	0.14	\\
N3211	&	\nodata	&	\nodata	&	2.68	&	0.44	\\
N3242	&	1.19	&	0.13	&	1.39	&	0.23	\\
N3918	&	\nodata	&	\nodata	&	1.37	&	0.22	\\
N5315	&	1.17	&	0.13	&	1.00	&	0.16	\\
N5882	&	1.21	&	0.14	&	1.04	&	0.17	\\
N6369	&	\nodata	&	\nodata	&	1.01	&	0.16	\\
N6445	&	\nodata	&	\nodata	&	1.23	&	0.18	\\
N6537	&	\nodata	&	\nodata	&	1.45	&	0.24	\\
N6563	&	\nodata	&	\nodata	&	1.07	&	0.15	\\
N6567	&	\nodata	&	\nodata	&	1.01	&	0.17	\\
N6572	&	\nodata	&	\nodata	&	1.00	&	0.16	\\
N6629	&	\nodata	&	\nodata	&	1.01	&	0.16	\\
N6751	&	\nodata	&	\nodata	&	1.00	&	0.15	\\
N6804	&	\nodata	&	\nodata	&	3.65	&	0.60	\\
N6826	&	1.15	&	0.12	&	1.00	&	0.16	\\
N6894	&	\nodata	&	\nodata	&	1.07	&	0.15	\\
N7008	&	\nodata	&	\nodata	&	1.53	&	0.25	\\
N7009	&	1.22	&	0.14	&	1.05	&	0.17	\\
N7027	&	\nodata	&	\nodata	&	1.41	&	0.23	\\
N7293	&	0.96	&	0.08	&	1.04	&	0.16	\\
N7354	&	\nodata	&	\nodata	&	1.25	&	0.21	\\
N7662	&	1.19	&	0.13	&	1.48	&	0.24	\\
\enddata
\tablenotetext{a}{ICFs for C/O and O/H were determined using the formulae in \citet{delgado14}. For He/H and N/O, we assumed that He$^{++}$/H$^+$+He$^+$/H$^+$=He/H and N$+$/O$^+$=N/O, respectively.}
\end{deluxetable}

\begin{deluxetable}{lccccccccc}
\tablecolumns{10}
\tablewidth{0pc}
\tabletypesize{\scriptsize}
\tablecaption{Total Elemental Abundances\label{abuns}}
\tablehead{
\colhead{PN}&	{M$_{ave}$/M$_{\odot}$}&{He/H}	&{$\sigma$(He/H)}&{C/O}&{$\sigma$(C/O)}&{N/O}&{$\sigma$(N/O)}&{O/H}&{$\sigma$(O/H)}
}

\startdata
FG1	&	0.6	&	0.13	&	0.01	&	\nodata	&	\nodata	&	0.61	&	0.19	&	3.07E-04	&	5.60E-05	\\
IC2149	&	1.2	&	0.11	&	0.01	&	\nodata	&	\nodata	&	0.14	&	0.04	&	2.20E-04	&	3.89E-05	\\
IC2165	&	2.3	&	0.11	&	0.01	&	1.37	&	0.18	&	0.36	&	0.06	&	2.23E-04	&	3.66E-05	\\
IC3568	&	2.5	&	0.12	&	0.01	&	0.65	&	0.14	&	0.04	&	0.03	&	2.96E-04	&	5.18E-05	\\
IC418	&	1.4	&	0.07	&	0.01	&	3.34	&	1.99	&	0.45	&	0.14	&	1.53E-04	&	2.74E-05	\\
IC4593	&	0.7	&	0.10	&	0.01	&	2.43	&	0.45	&	0.08	&	0.05	&	4.16E-04	&	7.65E-05	\\
N1501	&	1.7	&	0.12	&	0.01	&	\nodata	&	\nodata	&	0.20	&	0.04	&	4.15E-04	&	7.52E-05	\\
N2371	&	0.6	&	0.11	&	0.01	&	\nodata	&	\nodata	&	0.56	&	0.35	&	4.47E-04	&	9.04E-05	\\
N2392	&	1.6	&	0.08	&	0.01	&	1.06	&	0.54	&	0.05	&	0.05	&	1.68E-04	&	3.35E-05	\\
N2438	&	1.8	&	0.10	&	0.02	&	\nodata	&	\nodata	&	0.59	&	0.35	&	3.72E-04	&	7.27E-05	\\
N2440	&	2.8	&	0.13	&	0.01	&	0.99	&	0.14	&	1.71	&	0.61	&	3.10E-04	&	5.44E-05	\\
N2792	&	1.2	&	0.11	&	0.01	&	\nodata	&	\nodata	&	0.24	&	0.23	&	4.50E-04	&	9.81E-05	\\
N3195	&	2.1	&	0.14	&	0.01	&	\nodata	&	\nodata	&	0.64	&	0.21	&	4.78E-04	&	8.57E-05	\\
N3211	&	2.5	&	0.12	&	0.01	&	\nodata	&	\nodata	&	0.28	&	0.29	&	5.08E-04	&	1.09E-04	\\
N3242	&	1.2	&	0.11	&	0.01	&	0.82	&	0.10	&	0.10	&	0.03	&	3.40E-04	&	5.64E-05	\\
N3918	&	2	&	0.11	&	0.01	&	\nodata	&	\nodata	&	0.51	&	0.18	&	3.85E-04	&	7.68E-05	\\
N5315	&	2.4	&	0.13	&	0.02	&	0.75	&	0.09	&	1.58	&	0.45	&	3.59E-04	&	6.11E-05	\\
N5882	&	1.1	&	0.11	&	0.01	&	0.23	&	0.07	&	0.34	&	0.06	&	4.32E-04	&	7.81E-05	\\
N6369	&	3	&	0.13	&	0.01	&	\nodata	&	\nodata	&	0.50	&	0.14	&	4.98E-04	&	1.06E-04	\\
N6445	&	2.3	&	0.14	&	0.01	&	\nodata	&	\nodata	&	0.87	&	0.32	&	5.73E-04	&	1.06E-04	\\
N6537	&	3.7	&	0.17	&	0.02	&	\nodata	&	\nodata	&	8.84	&	3.47	&	1.65E-04	&	3.67E-05	\\
N6563	&	1.7	&	0.13	&	0.01	&	\nodata	&	\nodata	&	0.44	&	0.17	&	4.32E-04	&	8.17E-05	\\
N6567	&	0.7	&	0.10	&	0.02	&	\nodata	&	\nodata	&	0.31	&	0.13	&	2.28E-04	&	4.46E-05	\\
N6572	&	1.4	&	0.13	&	0.01	&	\nodata	&	\nodata	&	0.95	&	0.28	&	3.80E-04	&	7.20E-05	\\
N6629	&	1.6	&	0.11	&	0.01	&	\nodata	&	\nodata	&	0.15	&	0.08	&	4.10E-04	&	7.54E-05	\\
N6751	&	2.7	&	0.14	&	0.02	&	\nodata	&	\nodata	&	0.59	&	0.18	&	3.96E-04	&	7.03E-05	\\
N6804	&	1.4	&	0.11	&	0.01	&	\nodata	&	\nodata	&	0.12	&	0.10	&	4.01E-04	&	8.48E-05	\\
N6826	&	1.6	&	0.11	&	0.01	&	1.26	&	0.47	&	0.12	&	0.34	&	3.76E-04	&	6.96E-05	\\
N6894	&	1	&	0.13	&	0.01	&	\nodata	&	\nodata	&	0.79	&	0.27	&	3.81E-04	&	7.59E-05	\\
N7008	&	0.7	&	0.15	&	0.01	&	\nodata	&	\nodata	&	0.57	&	0.31	&	4.66E-04	&	8.99E-05	\\
N7009	&	1.4	&	0.12	&	0.02	&	1.74	&	1.16	&	0.41	&	0.11	&	5.10E-04	&	9.66E-05	\\
N7027	&	2.7	&	0.11	&	0.01	&	\nodata	&	\nodata	&	1.02	&	0.20	&	2.69E-04	&	5.63E-05	\\
N7293	&	2.1	&	0.12	&	0.01	&	0.52	&	0.91	&	0.74	&	0.59	&	4.31E-04	&	9.75E-05	\\
N7354	&	2.5	&	0.13	&	0.01	&	\nodata	&	\nodata	&	1.22	&	0.39	&	4.42E-04	&	8.51E-05	\\
N7662	&	1.2	&	0.12	&	0.01	&	0.78	&	0.13	&	0.12	&	0.03	&	2.95E-04	&	5.06E-05	\\
\enddata
\end{deluxetable}

Our final elemental abundances and uncertainties appear in Table~\ref{abuns}. Object names are provided in column~1, while column~2 contains our estimate of the progenitor star mass for that object. These masses were inferred according to the method described in the next subsection. Beginning with column~3, pairs of columns list the elemental number abundances and uncertainties for He/H, C/O, N/O and O/H. The uncertainties were rigorously determined by adding in quadrature the partial uncertainty contributions from each ion involved in the total element computation as well as the ICF uncertainty\footnote{The ICF uncertainty was unavailable in the case of N/O.}. The results provided in Table~\ref{abuns} will be analyzed in detail in \S\ref{results} following our detailed discussion of our method for determining progenitor masses.

\subsection{Progenitor Star Masses\label{progmass}}

Central star and progenitor masses were estimated by plotting the position of each central star in the log(L/L$_{\odot}$)-log~T$_{eff}$ plane along with theoretical post-AGB evolutionary tracks and interpolating between tracks for each of our 35 objects.  The values of log(L/L$_{\odot}$) and log~T$_{eff}$ were taken from \citet{frew08} for 32 of our 35 sample objects. For the three sample objects not included in \citet{frew08} (IC2165, IC3568 and NGC5315) we assumed the L and T values derived from models in \citet{henry15}.

We decided to base our analysis on the log(L/L$_{\odot}$) and log~T$_{eff}$ values for each of our objects found in \citet{frew08} because of the thoroughness of the procedures which he used to obtain these values.
In his compilation of log(L/L$_{\odot}$) values, Frew vetted all published V magnitude estimates for quality and then averaged the best values for each central star. Absolute visual magnitudes were then determined via a distance modulus, where distances were inferred from a new relation developed in \citet{frew08} between the H$\alpha$ surface brightness and nebular radius of a PN. Following the application of a bolometric correction, bolometric magnitudes were converted to solar luminosities. The effective temperature of each central star was determined by Frew using the  H and He Zanstra temperature methods in most cases. Table~\ref{masses} contains our adopted values for log(L/L$_{\odot}$) and log~T$_{eff}$ in columns~2 and 3, respectively, for each PN listed in column~1. 

\begin{deluxetable}{lccccc}
\tablecolumns{6}
\tablewidth{0pc}
\tabletypesize{\scriptsize}
\tablecaption{Progenitor Masses\label{masses}}
\tablehead{
\colhead{PN}&	{log(L/L$_{\odot}$)\tablenotemark{a}}	&	{log(T$_{eff}$)\tablenotemark{a}}	&	{M/M$_{\odot}$ (VW)\tablenotemark{b}}	&	{M/M$_{\odot}$ (MB)\tablenotemark{c}}	&	{M$_{ave}/M_{\odot}$}
}

\startdata
FG1	&	3.23	&	4.9	&	0.4	&	0.8	&	0.6	\\
IC2149	&	3.66	&	4.62	&	1.3	&	1.2	&	1.2	\\
IC2165	&	3.87	&	5.06	&	2.3	&	2.2	&	2.3	\\
IC3568	&	3.98	&	4.71	&	2.5	&	2.4	&	2.5	\\
IC418	&	3.72	&	4.58	&	1.5	&	1.3	&	1.4	\\
IC4593	&	3.41	&	4.6	&	0.5	&	0.9	&	0.7	\\
N1501	&	3.66	&	5.13	&	1.8	&	1.5	&	1.7	\\
N2371	&	2.98	&	5	&	0.5	&	0.8	&	0.6	\\
N2392	&	3.82	&	4.67	&	1.8	&	1.5	&	1.6	\\
N2438	&	2.31	&	5.09	&	2.0	&	1.5	&	1.8	\\
N2440	&	3.32	&	5.32	&	3.0	&	2.6	&	2.8	\\
N2792	&	3.18	&	5.1	&	1.3	&	1.2	&	1.2	\\
N3195	&	2.56	&	5.15	&	2.2	&	1.9	&	2.1	\\
N3211	&	2.76	&	5.21	&	2.6	&	2.3	&	2.5	\\
N3242	&	3.54	&	4.95	&	1.2	&	1.2	&	1.2	\\
N3918	&	3.7	&	5.18	&	2.0	&	2.0	&	2.0	\\
N5315	&	3.95	&	4.78	&	2.5	&	2.3	&	2.4	\\
N5882	&	3.52	&	4.83	&	1.0	&	1.1	&	1.1	\\
N6369	&	4.07	&	4.82	&	3.1	&	2.8	&	3.0	\\
N6445	&	2.97	&	5.23	&	2.4	&	2.1	&	2.3	\\
N6537	&	3.3	&	5.4	&	4.2	&	3.1	&	3.7	\\
N6563	&	2.34	&	5.09	&	1.9	&	1.4	&	1.7	\\
N6567	&	3.35	&	4.78	&	0.5	&	0.9	&	0.7	\\
N6572	&	3.72	&	4.84	&	1.5	&	1.3	&	1.4	\\
N6629	&	3.82	&	4.67	&	1.8	&	1.5	&	1.6	\\
N6751	&	3.97	&	5.02	&	2.7	&	2.6	&	2.7	\\
N6804	&	3.71	&	4.93	&	1.6	&	1.3	&	1.4	\\
N6826	&	3.81	&	4.7	&	1.7	&	1.5	&	1.6	\\
N6894	&	2.23	&	5	&	0.9	&	1.0	&	1.0	\\
N7008	&	3.12	&	4.99	&	0.5	&	0.8	&	0.7	\\
N7009	&	3.67	&	4.94	&	1.5	&	1.2	&	1.4	\\
N7027	&	3.87	&	5.24	&	2.8	&	2.6	&	2.7	\\
N7293	&	1.95	&	5.04	&	2.2	&	2.0	&	2.1	\\
N7354	&	3.95	&	4.98	&	2.5	&	2.5	&	2.5	\\
N7662	&	3.42	&	5.05	&	1.2	&	1.2	&	1.2	\\
\enddata
\tablenotetext{a}{All values for luminosity and effective temperature were taken directly from \citet{frew08} except in the cases of IC2165, IC3568 and NGC5315.  For those three objects, observed values listed in Tables~5 and 6 in \citet{henry15} were assumed.}
\tablenotetext{b}{Masses determined using the post-AGB tracks by \citet{vw94}}
\tablenotetext{c}{Masses determined using the post-AGB tracks by \citet{mb16}}
\end{deluxetable}

We experimented with two sets of post-AGB evolutionary tracks: those by \citet[Z=0.016, VW]{vw94} and \citet[Z=0.010, MB]{mb16}. Model sets differing in authorship as well as metallicity were chosen deliberately in order to test the effect upon inferred masses.  For each set we plotted tracks in a separate log(L/L$_{\odot}$)-log~T$_{eff}$ diagram and then placed our sample objects in the graph using our adopted values of these two stellar properties listed in Table~\ref{masses}.  

\begin{figure}
   \includegraphics[width=6in,angle=270]{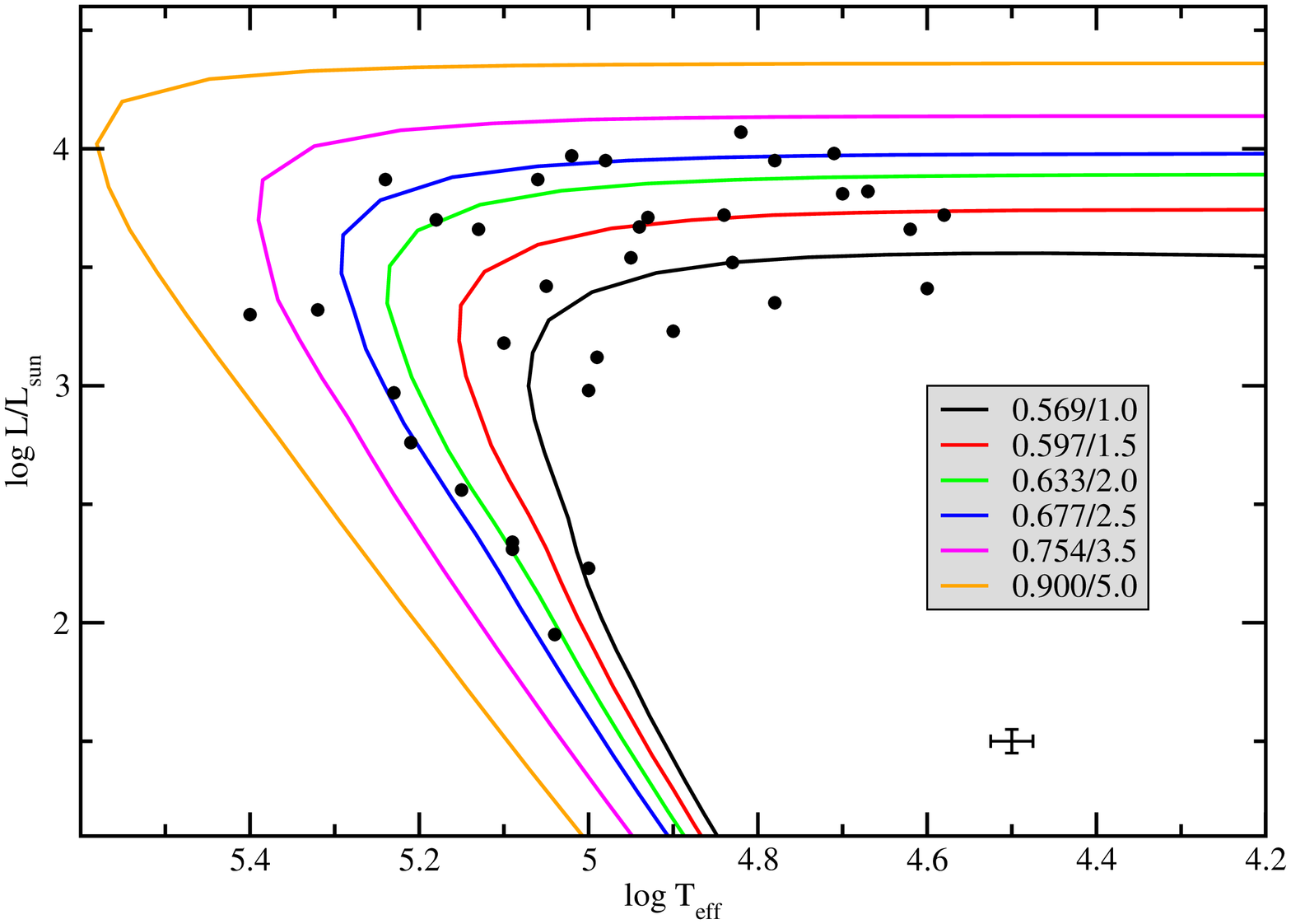} 
   \caption{log L/L$_{\odot}$ versus log T$_{eff}$. Solid colored lines show the post-AGB tracks of \citet{vw94} for Z=0.016. The legend indicates the correspondence between line color and remnant/progenitor mass. The positions of our 35 objects are shown with filled circles. The representative error bars located in the lower right are taken from Fig. 9.8 of \citet{frew08}.}
\label{fvw94}
\end{figure}

\begin{figure}
   \includegraphics[width=6in,angle=270]{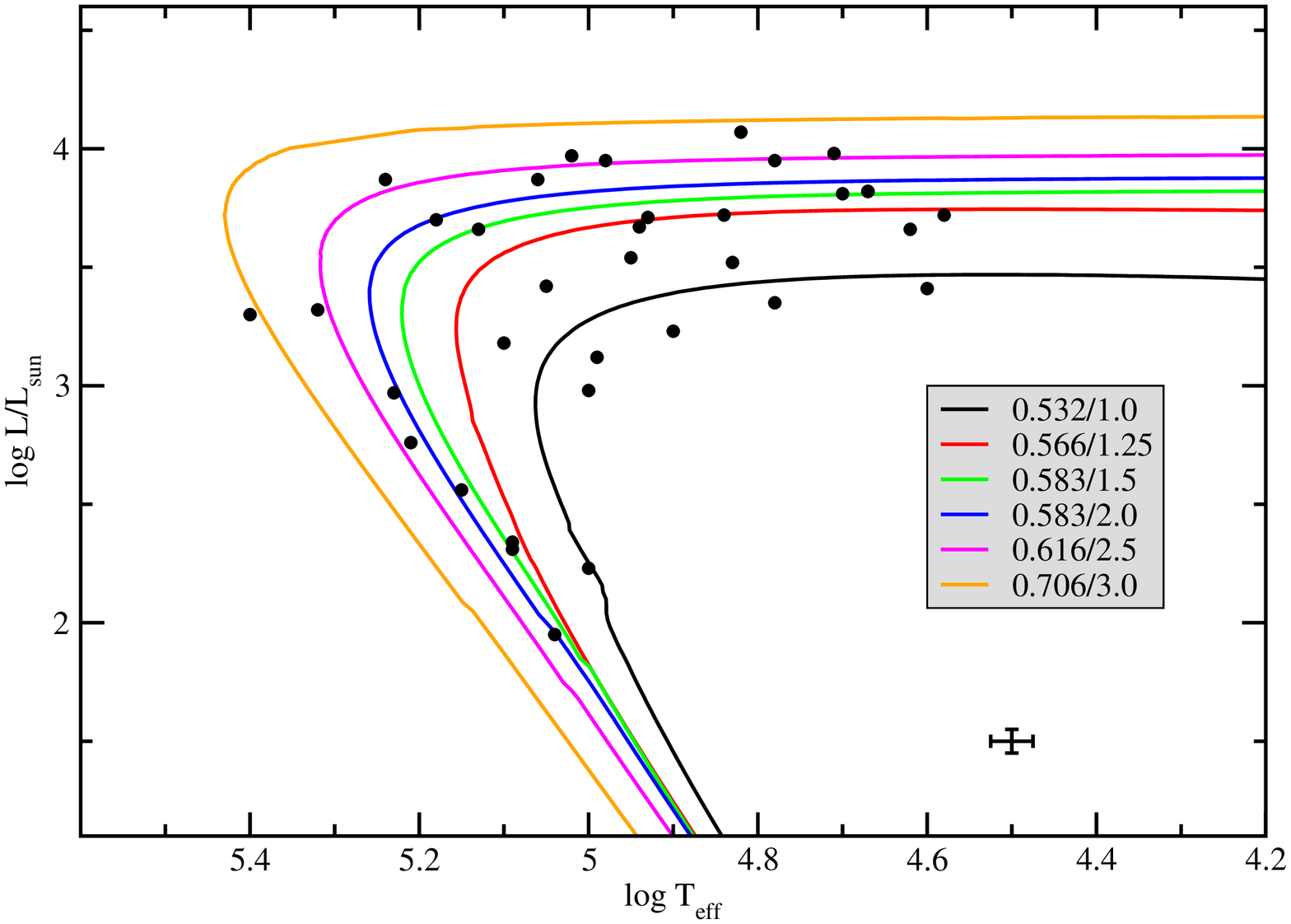} 
   \caption{log L/L$_{\odot}$ versus log T$_{eff}$. Solid colored lines show the post-AGB tracks of \citet{mb16} for Z=0.010. The legend indicates the correspondence between line color and remnant/progenitor mass. The positions of our 35 objects are shown with filled circles. The representative error bars located in the lower right are taken from Fig. 9.8 of \citet{frew08}.}
\label{fmb16}
\end{figure}

Figures~\ref{fvw94} and \ref{fmb16} show the positions of our sample objects in a log(L/L$_{\odot}$)-log~T$_{eff}$ plane along with the model tracks of VW and MB, respectively. 
 The final/initial mass associated with each track is designated by track color as defined in each figure's legend.  Representative error bars for the observed values, shown in the lower right of each figure, are taken directly from Fig.~9.8 of \citet{frew08}, since uncertainties for individual objects were not provided. Because each track is associated with a specific initial and final mass, we carefully measured each object's displacement from adjacent tracks and interpolated to find the mass values. The resulting initial masses determined in Figs.~\ref{fvw94} and \ref{fmb16} are listed in columns~4 and 5 of Table~\ref{masses}, respectively. The average of these two masses is listed in column~6 of that table as well as in column~2 of Table~\ref{abuns}.

\begin{figure}
   \includegraphics[width=6in,angle=270]{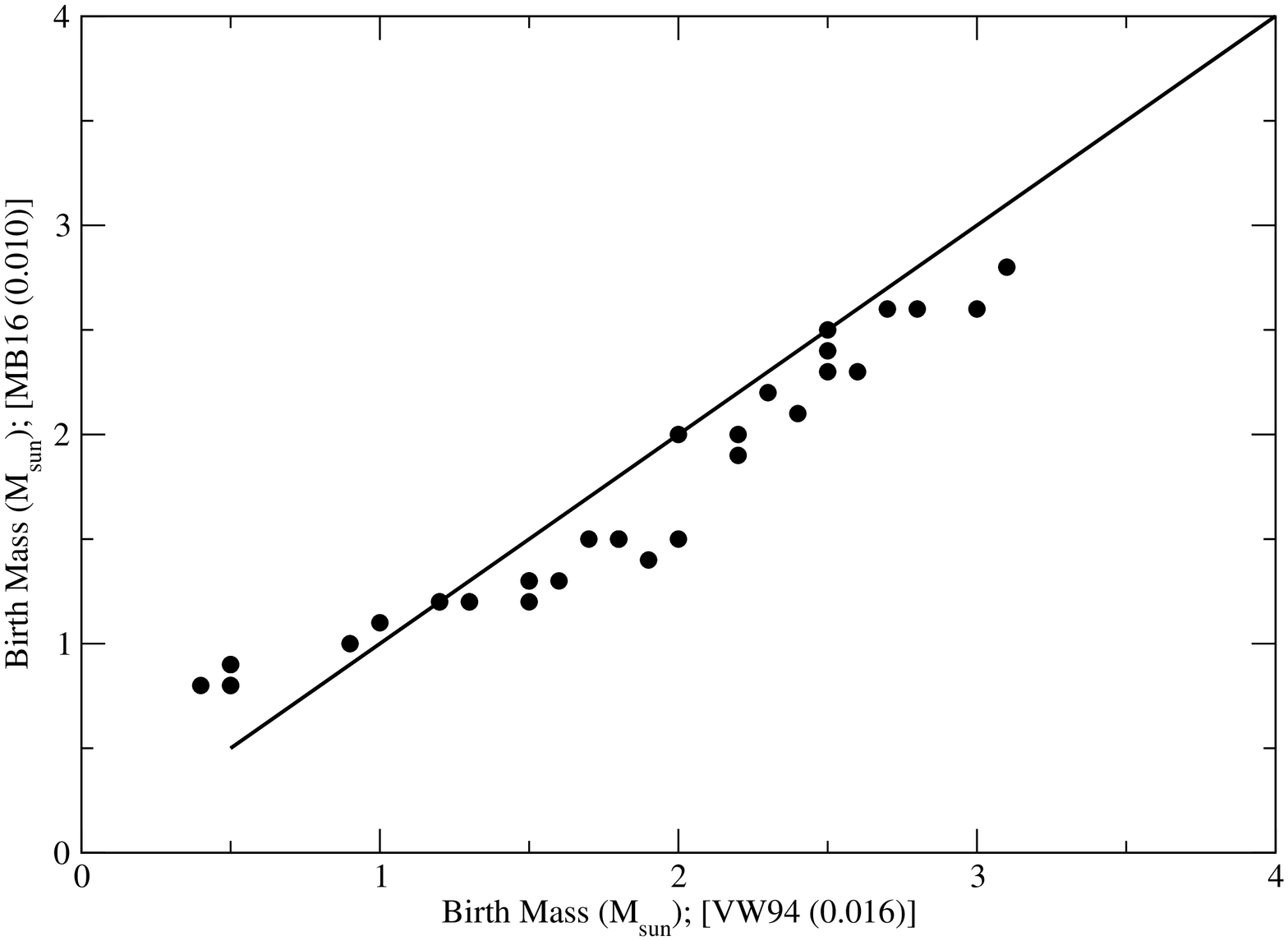} 
   \caption{Comparison of progenitor masses of our sample objects derived from the post-AGB tracks of \citet{mb16} (Fig.~\ref{fmb16}; vertical axis) and \citet{vw94} (Fig.~\ref{fvw94}; horizontal axis). The solid line indicates the one-to-one correspondence.}
\label{vwmb}
\end{figure}

Figure~\ref{vwmb} is a plot of masses from column~5 versus those in column~4 of Table~\ref{masses}. 
The straight line shows the one-to-one relation. For a vast majority of objects, the progenitor masses ($M_i$) determined using the MB tracks tend to be smaller than those determined from the VW tracks by about 0.3~M$_{\odot}$. This systematic difference is a direct consequence of the higher luminosity of the MB models during the constant luminosity stage resulting from the updated treatment of the evolutionary stages that precede the post-AGB stage. However, this offset is less than our estimated uncertainty of $\pm$0.5~M$_{\odot}$ and therefore is likely insignificant for our purposes here. Interesting exceptions are the five objects with $M_i\lesssim 1 M_\odot$, for which the MB tracks are slightly less luminous than those of VW. This leads to significantly higher extrapolated masses ($M_i\sim 0.8 M_\odot$) for three of the objects when using the MB tracks, instead of the unrealistically low $M_i\sim 0.5 M_\odot$ obtained with the VW tracks.

\section{Results \& Discussion\label{results}}

We now present a comparison of observed abundance ratios for our sample objects to several sets of theoretical model predictions of PN abundances in Figures~\ref{he2h}-\ref{c2ovn2o}. 
Note that in these figures, model tracks differing in metallicity but produced by the same code share the same line color. The metallicity of each model follows the code name in the legend and generally increases in value from solid to dashed to dotted line types. Solar abundance values from \citet{asplund09} are shown with black dotted lines. 

 To understand the differences in the predictions of
  the different theoretical models, and also to extract some physical
  insight from their comparison with the observations, it is necessary
  to keep in mind the different physical assumptions of each grid.
  The evolution of the surface abundances of AGB stellar models is
  particularly sensitive to the adopted physics on the AGB. In
  addition, the properties of the stellar models in advanced
  evolutionary stages, such as the AGB, are affected by the modeling
  of previous evolutionary stages. The latter is particularly true
  regarding the treatment of mixing processes such as rotationally
  induced mixing or convective boundary mixing (or overshooting)
  during H- and He-core burning stages. 
  
  \subsection{Description of the Model Codes}
  
  We now briefly review the
  treatment of these key ingredients in the four grids adopted here
  for the comparison: the MONASH grid
  \citep{2014MNRAS.445..347K, karakas16}, the LPCODE grid
  \citep{mb16}, the ATON grid \citep{ventura15, 2016MNRAS.462..395D}
  and the FRUITY database \citep{cristallo11,cristallo15}.  While all
  the models discussed here include an up-to-date treatment of the
  microphysics, and all of them neglect the impact of rotation, the
  theoretical models discussed in this section have some key
  differences in the modeling of winds and convective boundary mixing
  processes. These differences will affect the predicted evolution and
  final abundances during the TP-AGB. 
  
  Based on the treatment of winds
  on the AGB, the models can be roughly divided in two groups. On one
  hand we have the MONASH and FRUITY models that adopt a single relation between the pulsational period $P$ and the
  mass loss rate $\dot{M}$ for both C-rich and O-rich AGB stars. The
  mass loss recipe $\dot{M}(P)$ adopted by the MONASH models
  is the well know formula by \citet[eqs. 1, 2, \& 5]{1993ApJ...413..641V}, while the
  FRUITY models adopt a similar prescription derived by
  \citet[see their \S5]{2006NuPhA.777..311S}. On the other hand, we have the
  implementations by the ATON and LPCODE grids that incorporate a
  different treatment for the C-rich and the O-rich AGB winds. The
  ATON code adopts the empirical law by \citet[eqs. 1 \& 16 with $\eta_R=0.02$]{1995A&A...297..727B}
  reduced by a factor 50 for the O-rich phase and the theoretical mass
  loss rates by \citet[eqs. 1, 2 \& 3]{2008A&A...486..497W} for C-rich winds.  The LPCODE models adopt
  the empirical law by \citet{1998MNRAS.293...18G} for the C-rich
  phase while winds for the O-rich phase mostly follow the
  \citet{2005ApJ...630L..73S} law. These laws appear as eqs. 1, 2, 3, \& 5 in \citet{mb16}. 
  
  Even more important than the
  treatment of winds is the treatment of convective boundary mixing
  (or overshooting) during the TP-AGB phase as well as in previous
  evolutionary stages. Again the models can be roughly separated into
  two groups regarding the treatment of overshooting during
  core-burning stages. As before, on the one hand we have the
  MONASH and FRUITY
  models that do not include any kind of convective boundary mixing
  processes on the upper main sequence where stars have convective
  cores. However, later during the He-core burning stage, FRUITY models include
  convective boundary mixing in the form of semiconvection
  \citep{cristallo11}. And while the MONASH models do not include
  any explicit prescription for convective boundary mixing, a
  similar result would be expected from their adopted numerical
  algorithm to search for a neutrally stable point at the outer
  boundary of the convective core \citep{1986ApJ...311..708L}. On the
  other hand, the ATON and LPCODE models include overshooting
  on top of the H-burning core with its extension calibrated to fit
  the width of the upper main sequence. Both grids keep the same
  calibrated overshooting for the convective core during the core
  He-burning stage. From this difference alone in the treatment of
  convective boundary mixing before the TP-AGB, {\it one should expect
  third dredge up (TDU) and hot bottom burning (HBB) to develop at
  lower initial masses ($M_i$) in the ATON and LPCODE models than in
  the models of the MONASH and FRUITY grids}.  
  
  Regarding
  convective boundary mixing on the TP-AGB, two convective boundaries
  are key for the strength of TDU events during the TP-AGB [see
  \citet{2000A&A...360..952H}]. These are the boundary mixing at the
  bottom of the pulse drive convective zone (PDCZ) that develops in
  the intershell region during the thermal pulses, and the boundary
  mixing at the bottom of the convective envelope (CE). The inclusion
  of overshooting at both convective boundaries increases the
  efficiency of TDU and lowers the threshold in initial stellar mass
  above which TDU develops. In addition, the inclusion of overshooting
  at the bottom of the PDCZ leads to the dredging up of O from the CO
  core, increasing the intershell and surface O abundances.  
  
  The
  treatment of these convective boundaries varies widely in the four
  grids discussed here.  The MONASH models do not include any
  explicit prescription for convective boundary mixing. However, some
  overshooting at convective boundaries does occur as a
  consequence of the adoption of the numerical algorithm for the
  determination of the convective boundaries
  \citep{1986ApJ...311..708L}. On the contrary the FRUITY, ATON and
  LPCODE models adopt different implementations of an exponentially
  decaying mixing coefficient \citep{1996A&A...313..497F} beyond the
  formally convective boundaries and with different intensities. While
  FRUITY models include strong overshooting at the bottom of the CE
  but no overeshooting at the PDCZ, LPCODE models adopt a moderate
  overshooting at the base of the PDCZ and no overshooting at the
  bottom of the CE. Finally, the ATON models adopt a very small
  amount of overshooting both at the bottom of the PDCZ and the CE.
  
While there are strong arguments in favour of the inclusion of moderate
overshooting during the main sequence \citep{schaller92, pietrinferni04, weiss09,2012A&A...537A.146E} the situation
on the AGB is much less clear. In fact, trying to fit all available
observational constraints by means of a simple overshooting prescription
might not be even possible [see \citet{weiss09, karakas14, mb16}]. This fact, together with the lack of
compelling theoretical arguments and the lack of a common observational
benchmark for AGB theoretical evolution models has led authors to the
adoption of very different approaches.
 
  
  Finally we note that convection in the ATON code is computed with the full
  spectrum of turbulence convection, which leads to stronger HBB
  \citep{2005A&A...431..279V} when compared with models that adopt the
  standard mixing length theory [\citet{cristallo11},
  \citet{karakas16} and \citet{mb16}]. 
  
  In summary, we can roughly divide the four grids into two main
  groups: 1)~the MONASH and FRUITY models that neglect convective
  boundary mixing during the main sequence, do not include
  overshooting in the PDCZ and adopt a single wind formula for both
  the C- and O-rich phases; and 2)~the ATON and LPCODE models which
  calibrate overshooting during core H-burning to the width of the
  main sequence, adopt the same overshooting for the core-He burning
  phase, include some overshooting at the bottom of the PDCZ, and
  adopt different wind prescriptions for the C- and O-rich
  phases. Note, however, that all grids adopt different
    treatments of convective boundary mixing during the TP-AGB.

 In addition to the differences in the adopted physics, there is another
 difference related to the point at which each sequence is
 terminated. Due to the several convergence problems experienced by
 stellar models at the end of the AGB, different authors choose to
 stop their sequences at some point before the end of the AGB, missing
 the last thermal pulse(s). Although the efficiency of third dredge up
 drops at the end of the AGB, some significant changes in the surface
 abundances can still happen in the last thermal pulses. This is
 because when the H-rich envelope mass has already been reduced by
 more than one order of magnitude, a much smaller amount of processed
 material needs to be dredged up to the surface to affect the final
 surface abundances. This is an important difference between the
 FRUITY, MONASH and ATON models that do not reach the post-AGB
 phase and the LPCODE grid models which are computed up until the white dwarf
 stage. LPCODE models show abundance variations due to the timing of
 the last AGB thermal pulse.
 
 \subsection{Analysis\label{analysis}}

\begin{figure}
   \includegraphics[width=6in,angle=270]{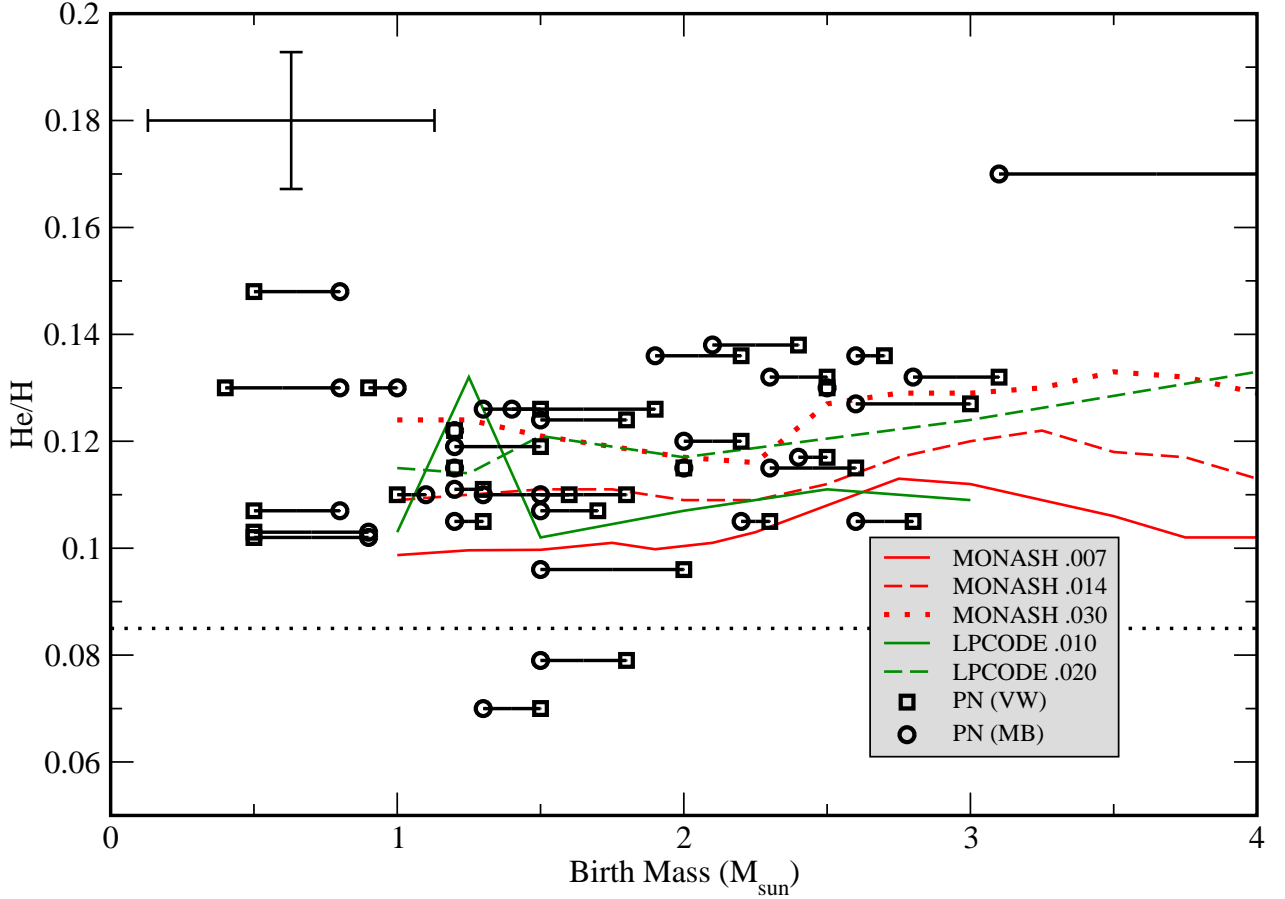} 
   \caption{He/H versus central star birth mass in solar units. PN are shown with connected pairs of open symbols. The squares represent objects whose progenitor masses were determined using the evolutionary tracks of \citet[Fig.~\ref{fvw94}]{vw94}, while the circles similarly refer to the tracks of \citet[Fig.~\ref{fmb16}]{mb16}. Error bars for individual objects have been suppressed for clarity, while a representative set of error bars is provided in the upper left corner of the plot. The horizontal black dotted line indicates the solar He/H value of 0.085 as determined by \citet{asplund09}.  Model predictions by the MONASH (red lines) and LPCODE (green lines) codes are shown for the metallicities given in the legend and designated in the graph by line type.}
\label{he2h}
\end{figure}

\begin{figure}
   \includegraphics[width=6in,angle=270]{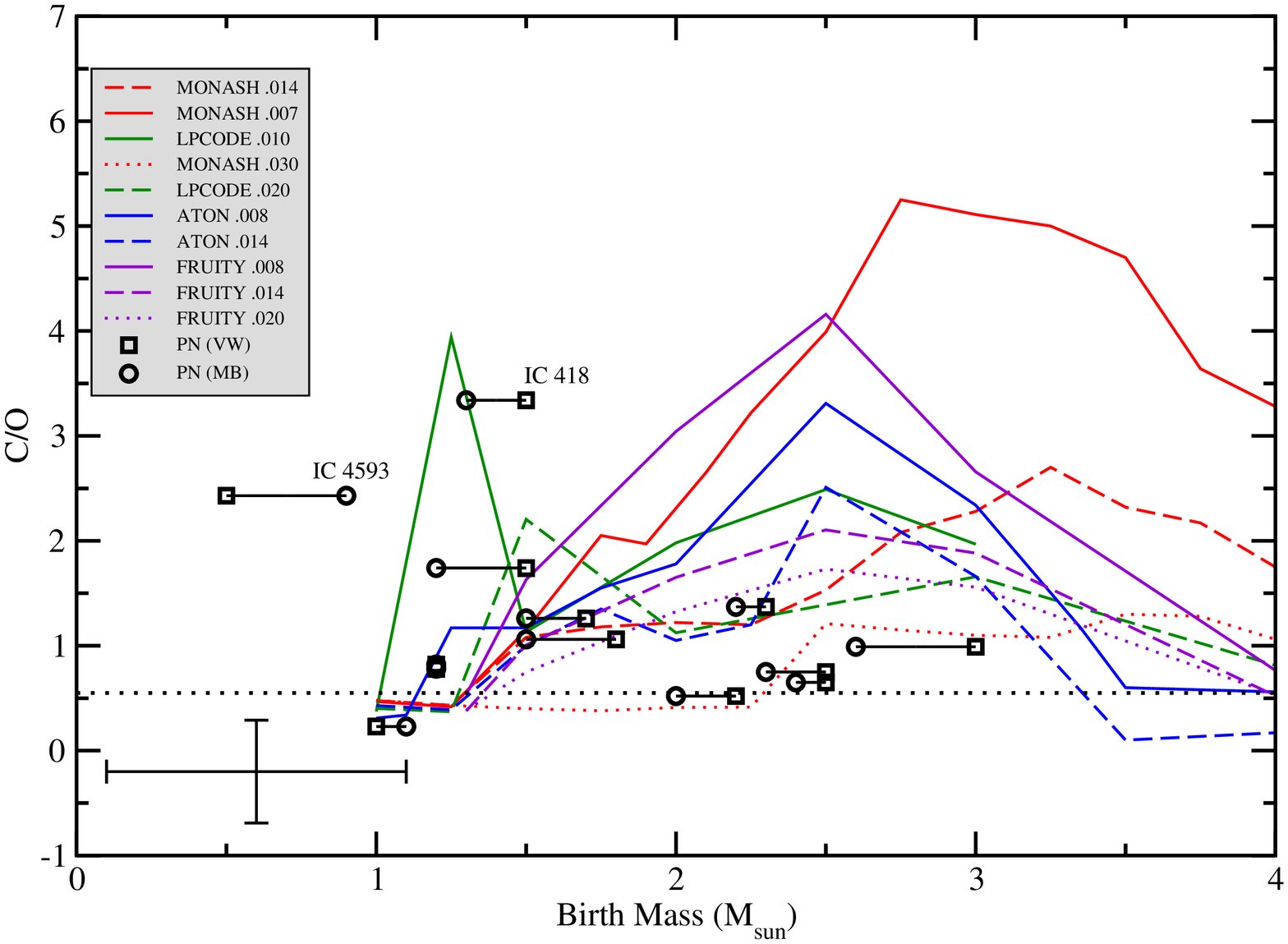} 
   \caption{C/O versus central star birth mass in solar units. PN are shown with connected pairs of open symbols. The squares represent objects whose progenitor masses were determined using the evolutionary tracks of \citet[Fig.~\ref{fvw94}]{vw94}, while the circles similarly refer to the tracks of \citet[Fig.~\ref{fmb16}]{mb16}. Error bars for individual objects have been suppressed for clarity, while a representative set of error bars is provided in the lower left corner of the plot.  The horizontal black dotted line indicates the solar C/O value of 0.55 as determined by \citet{asplund09}. Model predictions by the MONASH (red lines), LPCODE (green lines), ATON (blue lines) and FRUITY (violet lines) codes are shown for the metallicities given in the legend and designated in the graph by line type.}
\label{c2o}
\end{figure}

\begin{figure}
   \includegraphics[width=6in,angle=270]{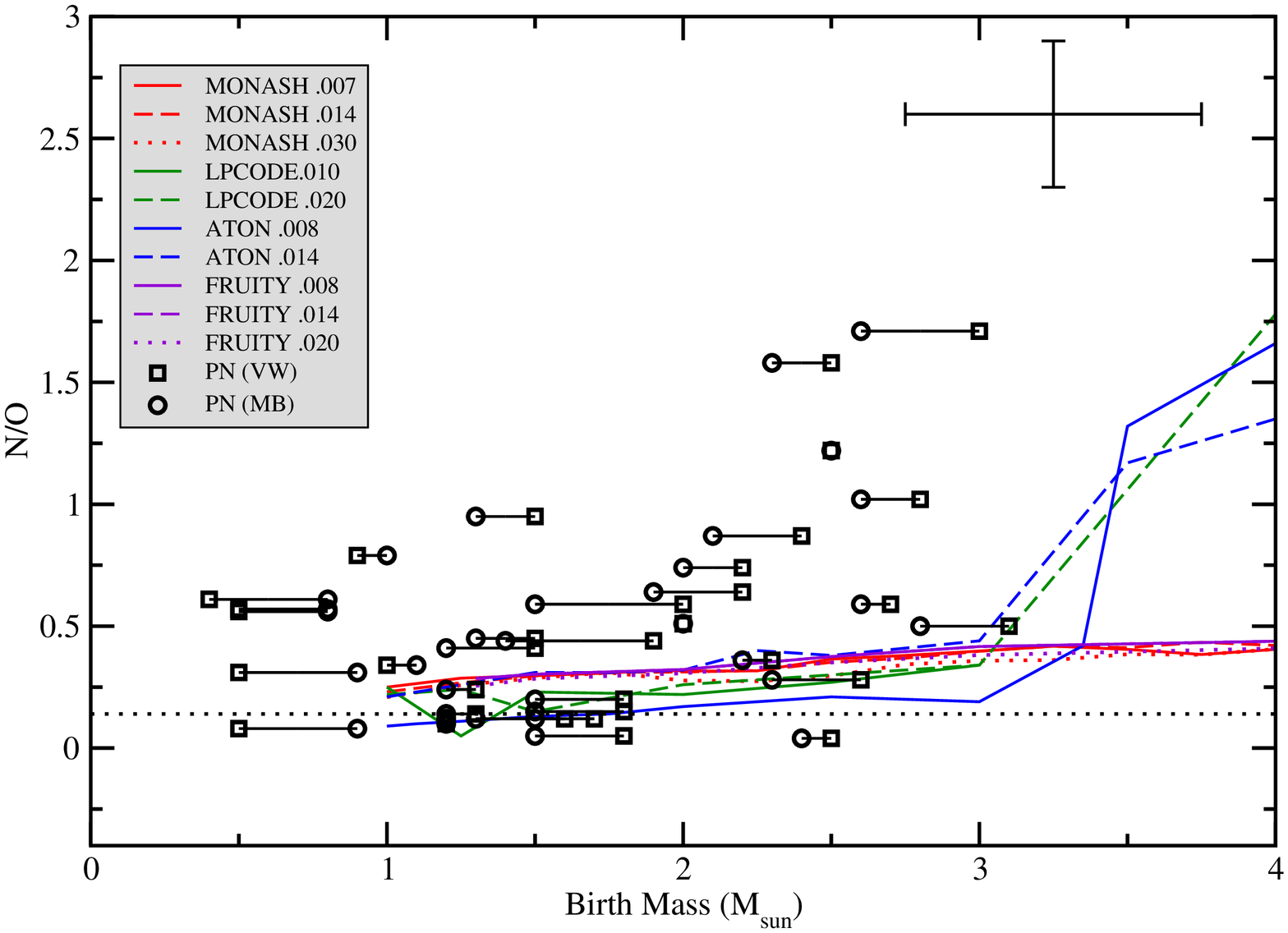} 
   \caption{N/O versus central star birth mass in solar units. PN are shown with connected pairs of open symbols. The squares represent objects whose progenitor masses were determined using the evolutionary tracks of \citet[Fig.~\ref{fvw94}]{vw94}, while the circles similarly refer to the tracks of \citet[Fig.~\ref{fmb16}]{mb16}. Error bars for individual objects have been suppressed for clarity, while a representative set of error bars is provided in the upper right corner of the plot.  The horizontal black dotted line indicates the solar N/O value of 0.14 as determined by \citet{asplund09}. Model predictions by the MONASH (red lines), LPCODE (green lines), ATON (blue lines) and FRUITY (violet lines) codes are shown for the metallicities given in the legend and designated in the graph by line type.}
\label{n2o}
\end{figure}

Our primary results involving the behavior of He/H, C/O and N/O versus progenitor mass appear in Figs.~\ref{he2h}-\ref{n2o}. 
Objects in our sample are shown with connected pairs of open squares and circles. The squares represent objects whose progenitor masses were determined using the evolutionary tracks of \citet[our Fig.~\ref{fvw94}]{vw94}, while the circles similarly refer to the tracks of \citet[our Fig.~\ref{fmb16}]{mb16}. Unpaired green circles represent objects for which the two derived masses were identical. For clarity, only a representative set of error bars is provided in each graph, where the vertical bar indicates the average of the relevant uncertainties given in Table~\ref{abuns}. Also included in the plots are model abundance predictions for PN ejecta by the MONASH, LPCODE, ATON and FRUITY grids (He/H predictions by the FRUITY and ATON grids were roughly constant at 0.10 and 0.095, respectively, and were not included in Fig.~\ref{he2h}). Line colors and types refer to the specific grid and metallicity, respectively, as defined in the figure legend. The horizontal and vertical black dotted lines show the solar values \citep{asplund09}.

The behavior of He/H versus progenitor mass is shown in
Fig.~\ref{he2h}. Relative to the solar value, all of our sample
members except two show He enrichment. 
Conspicuous outliers include NGC~6537
(He/H=0.17$\pm$.02) in the upper right and IC~418 (He/H=0.07$\pm$.01)
and NGC~2392 (He/H=0.08$\pm$.01) both located below the solar
line. NGC~6537 is a Peimbert Type~I PN, a class which
characteristically shows an enhanced He abundance.

Considering the He/H uncertainties, the MONASH and LPCODE model grids span the area occupied by the majority of points.  Note, though, that
in the case of the MONASH models, some of this success is achieved
only by including the Z=0.030 model set, i.e., a metallicity roughly
twice the solar value. This result is at odds with the metallicities
which we measured for our sample of objects, where nearly all have O/H
values\footnote{We note that oxygen may not be a reliable
  metallicity indicator if significant amounts of O are dredged up to
  the surface or destroyed by hot bottom burning during the TP-AGB as
  predicted by some models ---see section 3.1.1 in
  \cite{2016MNRAS.462..395D} and Table~3 in
  \citet{mb16}.}  in Table~\ref{abuns} at or below the solar level of $4.90
\times 10^{-4}$.  In addition both
the MONASH and LPCODE models predict a slight rise in He/H with
metallicity, but the observational uncertainties of He/H likely obscure this
theoretically predicted trend; if it indeed exists, it would be
difficult to see it in the data. And while deeper spectra may increase the S/N, accuracy would continue to be compromised due to the errors introduced by flux calibration, dereddening, instrumental effects and uncertainties associated with atomic constants, including collisional corrections. We feel that uncertainties of no less than $\pm$0.005 (a vertical error bar of 0.01) could likely be obtained.

In general the fact that most
measured He/H ratios are above the solar value is in line with the
expectations from stellar evolution theory, as all dredge up events
during post main sequence evolution lead to increases in the He/H
ratio.  It is well known that extra-mixing
processes are needed to explain the abundance patterns
in first red giant branch (RGB) stars located above the RGB bump
\citep{2007A&A...467L..15C,2010A&A...522A..10C,
2011A&A...533A.139W,lagarde12,maeder13}. We refer here to mixing processes in
addition to overshooting, such as rotationally induced mixing\footnote{Rotationally induced mixing includes different types of mixing processes
caused by the existence of rotation. These include mixing by meridional
circulation and  diffusion by shear  turbulence in differentially rotating
stars \citep{lagarde12,maeder13}.} or
thermohaline mixing\footnote{Thermohaline mixing  is a double diffusive process that can develop in
low-mass stars. This thermohaline instability takes place when the
stabilizing agent (heat) diffuses away faster than the destabilizing agent
(chemical composition), leading to a slow mixing process. Thermohaline
mixing can happen in low mass stars after the RGB-bump, and on the early
AGB \citep{lagarde12}, where an inversion of molecular weight is
created, by the 3He(3He,2p)4He reaction, on a dynamically stable structure.}. The fact that all grids fail to achieve the maximum observed values of He/H might be related to their neglect of extra-mixing processes on the pre-AGB evolution.



Figure~\ref{c2o} features the comparison of observations and models pertaining to C/O versus progenitor mass. It is interesting to note that C/O values are centered around 1.23 with a standard deviation of 0.85. Thus, despite the uncertainty in C/O indicated by the example error bar, the distribution of the 13 objects favors a supersolar value, the result of TDU. Both IC~418 (C/O=3.34$\pm$1.99) and IC~4593 (C/O=2.43$\pm$.45) exhibit C/O values which are at least twice the sample average. From our results in \S\ref{progmass} and Table~\ref{masses}, IC~418 had a progenitor mass of roughly 1.4$\pm$.5~M$_{\odot}$, while IC~4593's mass was originally 0.7$\pm$.5~M$_{\odot}$. The only model in Fig.~\ref{c2o} which predicts this much excess C within the mass range of the two progenitor stars is the one of $M_i=1.25 M_\odot$ and $Z=0.010$ in the LPCODE grid. Interestingly, that model attains its high surface carbon abundance due to a final thermal pulse when the mass of the central star is already reduced to $0.593 M_\odot$. In this circumstance TDU leads to the mixing of $M_{\rm TDU}\simeq 0.003 M_\odot$ from the H-free core into a H-rich envelope of $M^H_{\rm env}\simeq 0.027 M_\odot$, significantly increasing the surface carbon abundance of the star. This example shows why it is necessary to keep in mind that final AGB thermal pulses coupled with low envelope masses can significantly change the surface abundances from those predicted by AGB stellar evolution models which are not computed to the very end of the AGB. Yet, it is necessary to emphasize that if the mass ejected after the last thermal pulse is too small, the final abundances of the central stars might be different from those displayed by their surrounding PN. The nebula
might not be homogeneous and may be dominated by the material ejected
before the star altered its surface composition in the last thermal pulse.

Each of the four sets of model tracks displayed in Fig.~\ref{c2o} generally predicts two trends regarding C/O in PN: 1) as progenitor mass increases, C/O increases slowly, peaks around 2.5-3.0~M$_{\odot}$ and then decreases; and 2) for constant progenitor mass, C/O increases with decreasing metallicity. Both of these predicted trends are well-known and the presumed causes are nicely summarized in \citet[\S3.3]{karakas14}. In an AGB star, C is produced (and also dredged-up from the CO core) within the periodically unstable He shell by the triple alpha process and is subsequently transported to the H-rich outer envelope during TDU. According to models, the amount of C that is mixed up into the envelope is directly related to the efficiency of the dredge-up process, where the dredge-up efficiency is characterized by the ratio of the mass of material brought to the surface relative to the increase in mass of the C-O core during the process. Models indicate that this efficiency increases independently with increasing progenitor mass and decreasing metallicity. However, this process begins to be damped as the stellar mass approaches 4~M$_{\odot}$ in the case of the MONASH grid and 2.5-3~M$_{\odot}$ for the other three grids as C is converted to N via the CN cycle during HBB. The difference between the MONASH grid and the other three grids could be related to the lack of convective boundary mixing in the high mass models of the former grid, which leads to a less efficient HBB.

We turn now to the behavior of N/O versus progenitor mass featured in
Fig.~\ref{n2o}. Here we see that roughly 30\% of our objects exceed
the solar value of 0.14 for N/O by more than their uncertainties. We
also observe an upward trend in N/O in the data with increasing birth mass up to
about 3~M$_{\odot}$. 

The apparent nitrogen enrichment below
1.5~M$_{\odot}$ is likely the result of dredge-up events before the
AGB phase.  However, the upward trend beyond this point is rather substantial and likely is
the product of HBB. Interestingly, the lowest mass at which HBB is
predicted by the ATON and LPCODE models to begin is around
3~M$_{\odot}$ while MONASH and FRUITY models predict the onset of HBB at
around 5~M$_{\odot}$ (outside of the figure range). This difference is mostly due to the implementation of overshooting during the main sequence evolution in ATON and LPCODE models. 

Yet, the upward trend of N/O in our PN sample occurs at an even lower progenitor mass, with high N/O values corresponding to $M_i\gtrsim 2.25~M_{\odot}$. If our stellar mass determinations are reasonably correct, this result confirms the well established need to include overshooting in the modeling of the upper main sequence, and perhaps the need to include some additional mixing processes like rotation-induced mixing in main sequence intermediate mass stars \citep[Fig.~9]{2012A&A...537A.146E}. 

An additional shortcoming of the models is that
none of the sets spans the entire region occupied by our PN. {\it In particular, the
observations clearly suggest that stars with progenitor masses below
3~M$_{\odot}$ produce higher levels of N than are predicted by any of the models.}
As mentioned above, the failure of the models to account for the observed
  abundances of N in low-mass stars might be pointing to the need to
  include other mixing processes, such as rotation-induced mixing,
  during previous evolutionary stages \citep{2010A&A...522A..10C}.

\begin{figure}
   \includegraphics[width=6in,angle=270]{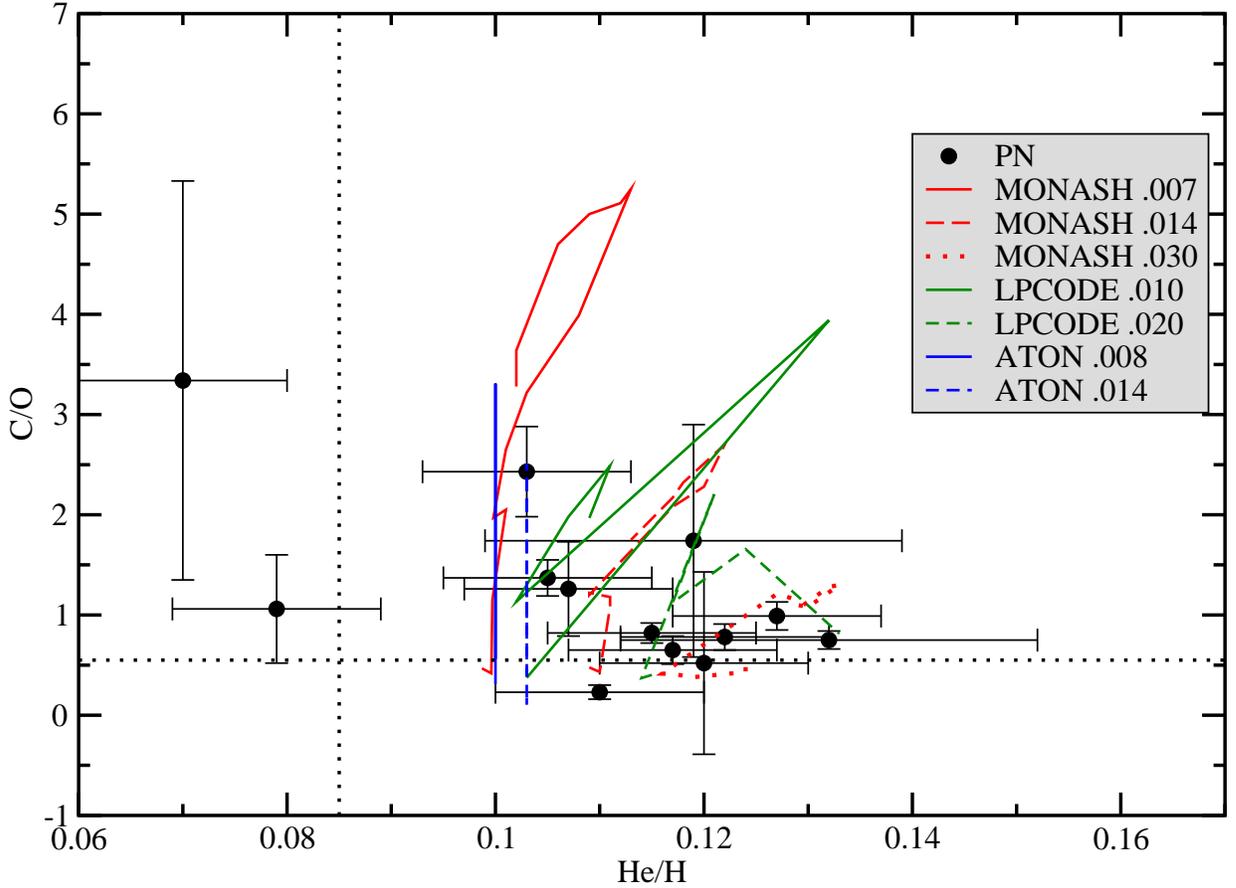} 
   \caption{C/O versus He/H. PN are shown with filled black circles. Solar ratios from \citet{asplund09} are shown with dotted black lines. Model predictions by the MONASH (red lines), LPCODE (green lines) and ATON (blue lines) codes are shown for the metallicities given in the legend and designated in the graph by line type. Note that the line for the ATON 0.014 model is purposely offset slightly to the right of the ATON 0.008 model to distinguish them, as otherwise they would lie on top of each other.}
\label{c2ovhe2h}
\end{figure}

\begin{figure}
   \includegraphics[width=6in,angle=270]{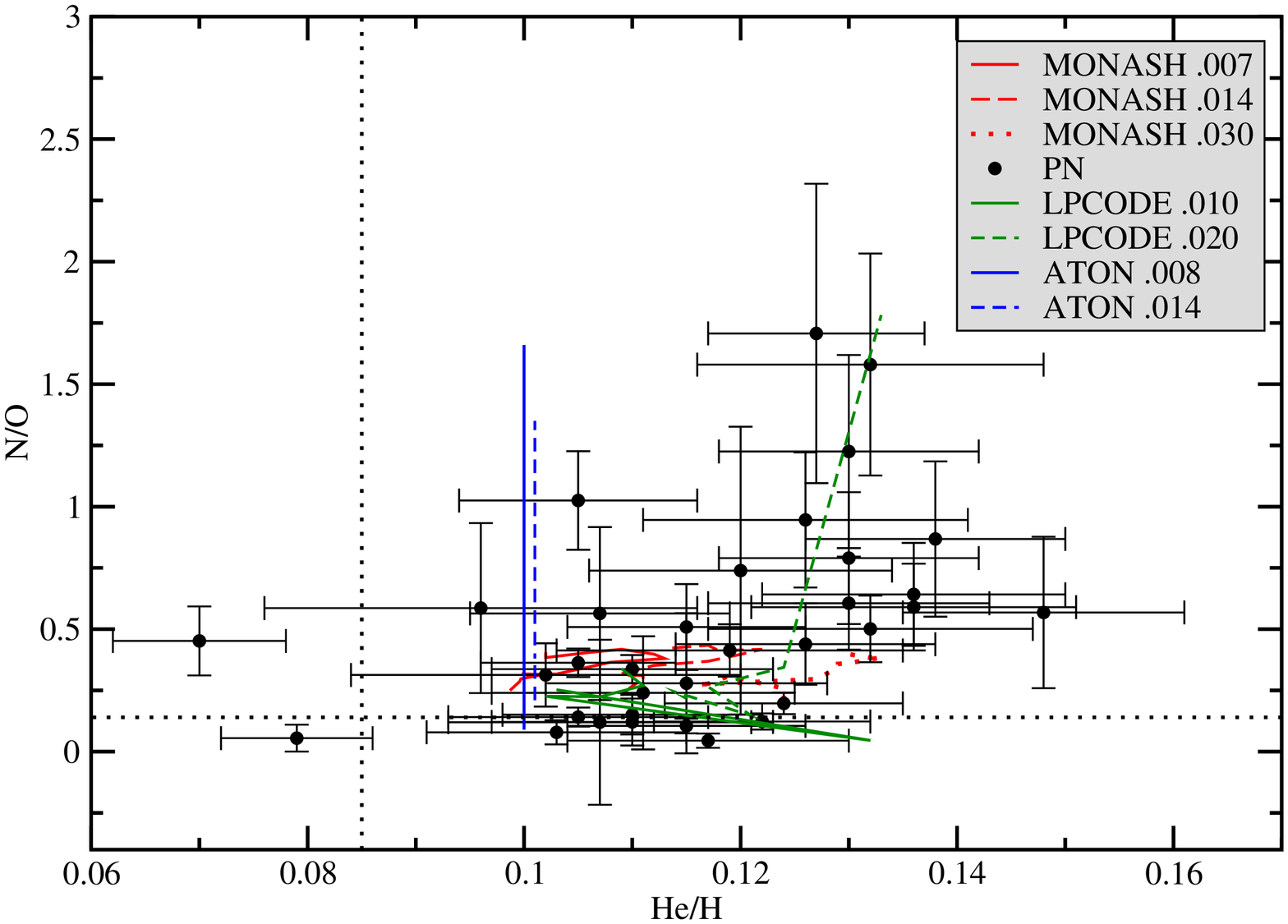} 
   \caption{N/O versus He/H. PN are shown with filled circles. Solar ratios from \citet{asplund09} are shown with dotted black lines. Model predictions by the MONASH (red lines), LPCODE (green lines) and ATON (blue lines) codes are shown for the metallicities given in the legend and designated in the graph by line type. Note that the line for the ATON 0.014 model is purposely offset slightly to the right of the ATON 0.008 model to distinguish them, as otherwise they would lie on top of each other.}
\label{n2ovhe2h}
\end{figure}

\begin{figure}
   \includegraphics[width=6in,angle=270]{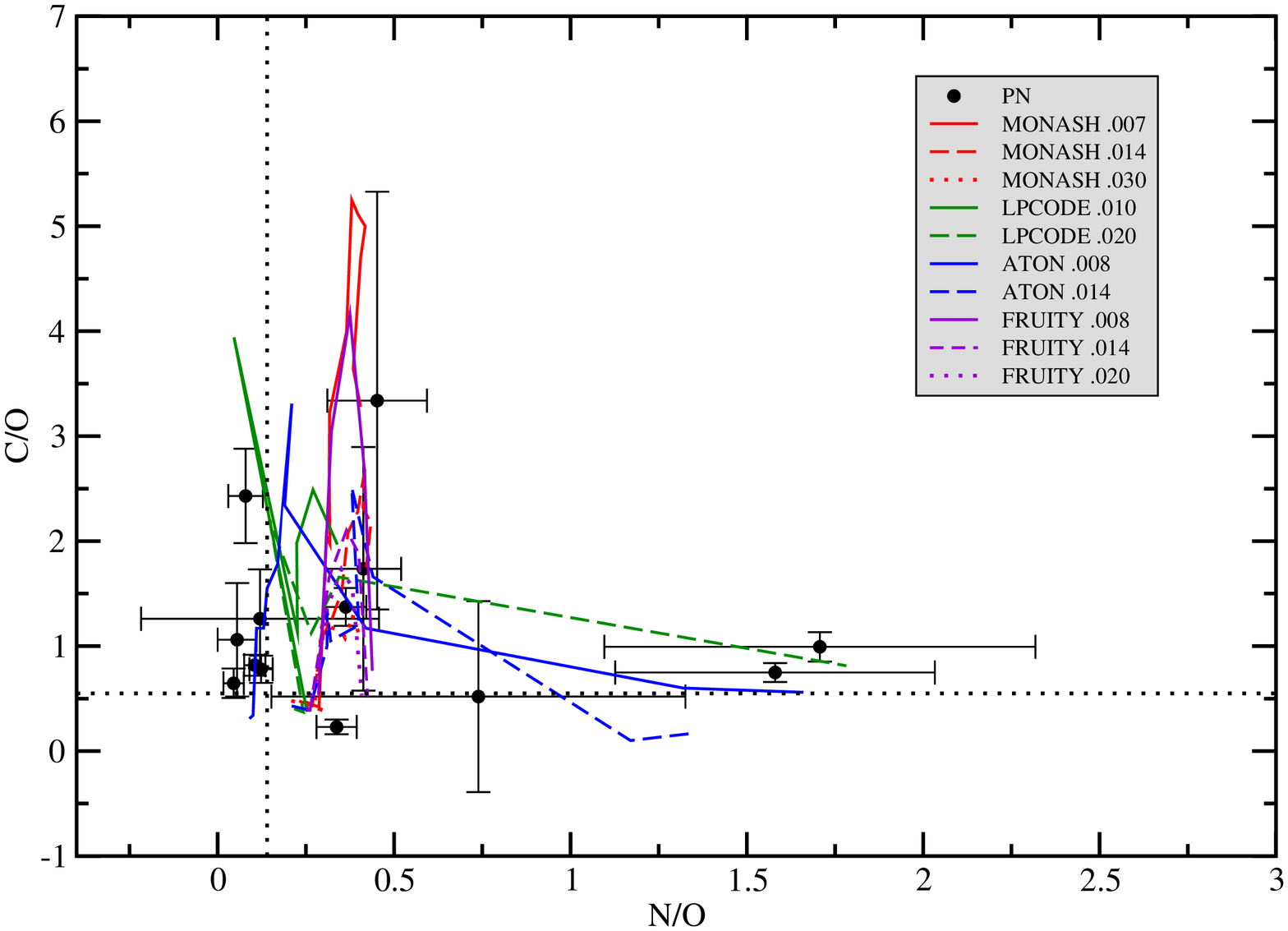} 
   \caption{C/O versus N/O. PN are shown with filled black circles. Solar ratios from \citet{asplund09} are shown with dotted black lines. Model predictions by the MONASH (red lines), LPCODE (green lines), ATON (blue lines) and FRUITY (violet lines) codes are shown for the metallicities given in the legend and designated in the graph by line type.}
\label{c2ovn2o}
\end{figure}

Figures \ref{c2ovhe2h}, \ref{n2ovhe2h} and \ref{c2ovn2o} compare observations and models in terms of one element ratio versus another one. Model tracks apply only to progenitor masses between 1 and 4 solar masses.

 Figure~\ref{c2ovhe2h} is a plot of C/O versus He/H, where we observe no apparent correlation between the values of these two ratios. As we saw earlier in Fig.~\ref{c2o}, the observed He/H ratio for all but IC~418 and NGC~2392 is above the solar value. This strongly suggests that a majority of the objects in our sample experienced significant He enrichment during their evolution. All of the ATON models within the 1-4 M$_{\odot}$ range predict a He/H value of 0.10, hence the straight vertical lines for those models. We have offset their track for the 0.014 metallicity models to the right slightly to help distinguish the two tracks. The model tracks of the MONASH and LPCODE grids are consistent with the observations in the sense that each model set spans the space occupied by the bulk of the sample objects, i.e., those 11 PN which have He/H$\ge$0.10. The ATON models appear to span the observed C/O values, but lack the range in He/H exhibited by the data. 
  
The observational data in Fig.~\ref{n2ovhe2h} suggest that the N enrichment seen earlier in Fig.~\ref{n2o}  may be coupled with He enrichment in the sense that large N/O values occur at high levels of He/H, although the large uncertainties in both N/O and He/H cloud the issue. Interestingly, a clear positive trend in N/O versus O/H was reported by \citet{kaler79} for Galactic disk PN, while \citet{kb94} observed similar behavior in Type~I PN only. The MONASH models fail to predict such behavior, as does the Z=0.01 LPCODE model track. This is due to the lack of efficient HBB in these models, as the MONASH models do not show HBB for $M_i<4M\odot$ and the LPCODE grid only reaches $M_i=3M\odot$ for this grid. On the contrary,  LPCODE models with Z=0.02  do show an upward trend in N/O as He/H increases due to the action of TDU and HBB during a large number of thermal pulses in the $M_i=4 M_\odot$ model. The ATON models seem to span the observed N/O values, although again there is no reported range in their He/H values. Overall, there is little theoretical evidence that any of the model grids completely spans the point positions of the observational data, a result we also see in Fig.~\ref{n2o}. We conclude that the possible observational trend in Fig.~\ref{n2ovhe2h} previously seen in Fig.~\ref{n2o} is likely reflecting the action of TDU and HBB.

Finally, Fig.~\ref{c2ovn2o} shows the relation of C/O versus N/O for the 13 objects for which we have C measurements. As we saw in Fig.~\ref{c2o}, these data exhibit a wide variation in the C/O ratio, with several objects having values significantly larger than the solar value. These same objects also have relatively low values of N/O, where ratios range from near solar to slightly above it. Then there are the three PN with solar C/O values that appear to be decidedly enriched with N. All model sets predict a significant variation in enhanced C/O at relatively low N/O, while at higher N/O levels the C/O values approach the solar value of 0.55. The data appear to be consistent with the models, and generally speaking, all model sets appear to span the empirical data sufficiently, although the high N/O region contains only three PN. The data in this figure are consistent with the theoretical expectation that C and N are anti-correlated, as C from TDU is subsequently destroyed during HBB to produce N.

Summarizing our detailed comparison of models and observations, the empirical trends seen in Figs. 9 and 12, and perhaps 11, suggest the existence of HBB in stars with birth masses less than 4~M$_{\odot}$, something that is only attained by models that include
overshooting on the main sequence (ATON, LPCODE). 

In more general terms, however, observations, when combined with model predictions of four independent model grids, currently demonstrate that all four grids are compatible with the data except in the case of N/O. That is, all grids seem capable of spanning the distribution of points in the cases of C/O and He/H. We suggest that future computational efforts consider the implication that the onset of HBB occurs at a lower initial mass than previously believed. This is the most important result of our study.

\section{Summary and Conclusions}

Helium, carbon and nitrogen are known through observations to be synthesized by stars within the mass range of 1-8~M$_{\odot}$ (low and intermediate mass stars, or LIMS). We demonstrated this plainly in Figs.~\ref{he2hvo2h_BIG}-\ref{n2ovo2h_BIG}, where we saw that the He/H, C/O and N/O abundance ratios as a function of metallicity in a large sample of PN systematically fall above ISM values for the same ratios measured by stars and H~II systems. 

To evaluate the significance of the relative contribution that LIMS make to the galactic chemical evolution of these three elements, we need to determine the amount of He, C and N that a star produces and releases into the interstellar medium, i.e., the stellar yield. Fortunately, a portion of this ejected matter forms a planetary nebula, and from the emission spectra produced by these objects, we are able to measure the abundances of He, C, N among other elements. Since theoretical models of LIMS predict both the total yield and the PN abundance, by comparing the observed abundances to theoretical predictions of the same we can simultaneously infer the yield.


The goal of this project has been to make a detailed comparison between observationally determined abundances of the elements He, C and N in planetary nebulae with theoretical predictions of the same by four different grids of stellar evolution models. We have carefully selected PN for which high quality spectra and good determinations of the luminosity and effective temperature of each associated central star are available. The optical and UV spectra consist exclusively of our own observations made with ground-based telescopes as well as HST/STIS and IUE. 

To ensure homogeneity, all spectral data were reduced and measured in a consistent manner, and abundances were all determined using the same algorithms. Central star luminosities and effective temperatures in all but three cases were taken from \citet{frew08}, and central and progenitor star masses were inferred by plotting these values in L-T diagrams containing evolutionary tracks from \citet{vw94} and \citet{mb16}. 

Our final sample contained 35 Galactic PN, 13 of which have C abundances measured from UV lines available. These 35 objects vary widely in morphology. All are either categorized as Peimbert type I or II. And most are located in the Galactic thin disk within 2~kpc of the Sun.

Combining the inferred abundances and stellar masses, we conclude the following:

\begin{enumerate}

\item {\it The mean values of N/O across the observed progenitor mass range of 1-3~M$_{\odot}$ are well above the solar value.} With respect to current theory, this is an unexpected result and suggests that extra-mixing is required in this stellar group to explain the N enrichment. Our results also suggest an increase in N/O with progenitor mass for M$>$2~M$_{\odot}$, implying that the onset of hot bottom burning occurs at lower masses than previously thought.

\item {\it All but two of our sample PN clearly show evidence of He enrichment relative to the solar value.} This is expected, since both first and third dredge-up mix He-rich material into the stellar atmosphere prior to PN formation from expelled atmospheric matter.

\item {\it The average value of measured C/O within our sample is 1.23, well above the solar value of 0.55 \citep{asplund09}.} The standard deviation for the sample is 0.85. Evidence of C enrichment is present in roughly half of the sample of 13 objects for which we measured the C abundance. Interestingly, the PN with the higher C/O values seem to come from low mass progenitors with $M\approx 1\ M_{\odot}$.


\item {\it The model grids to which we compared the observations successfully span the data points in the case of C/O. The models are also consistent with some, but not all, of the objects in terms of He/H. However, all of the models seem to fail in the case of N/O.}

\end{enumerate}


Our finding of elevated N/O in low mass stars, possibly due to an earlier-than-expected onset of HBB and/or the presence of extra-mixing, is the most significant result of our study. Further confirmation of this result will help markedly in the ongoing efforts to determine the provenance of N in the context of galactic chemical evolution. Because stars of masses between 1 and 3~M$_{\odot}$ are roughly five times more numerous than stars between 3 and 8~M$_{\odot}$ (assuming a simple Salpeter initial mass function), the potential impact of these low mass stars on the question of the chemical evolution of nitrogen is obviously significant.
  
\acknowledgments

The anonymous referee of our paper offered many helpful suggestions for improvement, and we thank him/her for performing such a careful review. We also thank Paolo Ventura and Sergio Cristallo for providing answers to our enquiries regarding the details of the ATON (Ventura) and FRUITY (Cristallo) model predictions and in some cases sending us additional output. We also appreciate the help provided by Gloria Delgado-Inglada concerning her group's recently updated ionization correction factors. Portions of the UV data employed in our project came from HST Program number GOÐ12600.  B.G.S. is grateful for summer support by the NSF through the Research Experience for Undergraduates
program. M.M.M.B. was supported by ANPCyT through grant PICT-2014-2708 and by a Return Fellowship from the Alexander von Humboldt Foundation. Finally, R.B.C.H., B.G.S., K.B.K., and B.B. are grateful to their home institutions for travel support.


\end{document}